\documentclass{article} 

\usepackage{iclr2027_conference,times}

\usepackage{amsmath}
\usepackage{amssymb}
\usepackage{hyperref}
\usepackage{url}
\usepackage{graphicx}
\usepackage{booktabs}
\usepackage{multirow}
\usepackage{array}
\usepackage{algorithm}
\usepackage{algorithmic}
\usepackage{caption}
\usepackage{longtable}
\usepackage{xcolor}
\usepackage{listings}

\newcolumntype{C}[1]{>{\centering\arraybackslash}p{#1}}

\definecolor{codegray}{gray}{0.95}
\definecolor{codekw}{rgb}{0.13,0.13,0.6}
\definecolor{codecomment}{rgb}{0.30,0.55,0.30}
\definecolor{codestring}{rgb}{0.65,0.15,0.15}
\lstdefinestyle{prompt}{%
  basicstyle=\ttfamily\scriptsize,
  breaklines=true,
  breakindent=0pt,
  columns=fullflexible,
  keepspaces=true,
  frame=single,
  rulecolor=\color{gray!50},
  backgroundcolor=\color{codegray},
  xleftmargin=4pt,xrightmargin=4pt,
  aboveskip=6pt,belowskip=6pt,
}
\lstdefinestyle{pycode}{%
  language=Python,
  basicstyle=\ttfamily\scriptsize,
  keywordstyle=\color{codekw}\bfseries,
  commentstyle=\color{codecomment}\itshape,
  stringstyle=\color{codestring},
  breaklines=true,
  showstringspaces=false,
  columns=fullflexible,
  keepspaces=true,
  frame=single,
  rulecolor=\color{gray!50},
  xleftmargin=4pt,xrightmargin=4pt,
  aboveskip=6pt,belowskip=6pt,
  morekeywords={triton,tl},
}

\newcommand{\calK}{\mathcal{K}}
\newcommand{\calP}{\mathcal{P}}
\newcommand{\calV}{\mathcal{V}}
\newcommand{\calA}{\mathcal{A}}
\newcommand{\calM}{\mathcal{M}}
\newcommand{\calB}{\mathcal{B}}
\newcommand{\calC}{\mathcal{C}}

\newtheorem{definition}{Definition}

\title{KernelOPT: Dispatch-Aware Agentic Search\\ for GPU Kernel Optimization}

\author{
Aheli Poddar\thanks{Equal contribution (joint first authors).}\thanks{Corresponding author.}\thanks{The first four authors completed this work during an internship with the PyTorch team at Red Hat.} \\
Red Hat \\
\texttt{ahpoddar@redhat.com} \\
\And
Sanskar Prasad\footnotemark[1]\footnotemark[3] \\
Red Hat \\
\texttt{sanspras@redhat.com} \\
\And
Arindam Samanta\footnotemark[3] \\
Red Hat \\
\texttt{arsamant@redhat.com} \\
\AND
Subha Chakraborty\footnotemark[3] \\
Red Hat \\
\texttt{subhchak@redhat.com} \\
\And
Vishal Goyal \\
Red Hat \\
\texttt{visgoyal@redhat.com} \\
\And
Rohit Singh Rathaur \\
Red Hat \\
\texttt{rrathaur@redhat.com} \\
}

\iclrfinalcopy            

\makeatletter
\renewcommand\maketitle{\par
  \begingroup
    \def\thefootnote{\fnsymbol{footnote}}%
    \def\@makefnmark{\hbox{\normalfont\textsuperscript{\@thefnmark}}}%
    \long\def\@makefntext##1{\parindent 1em\noindent
      \hbox to1.8em{\hss$\m@th^{\@thefnmark}$}##1}%
    \@maketitle \@thanks
  \endgroup
  \setcounter{footnote}{0}%
  \let\maketitle\relax \let\@maketitle\relax
  \gdef\@thanks{}\gdef\@author{}\gdef\@title{}\let\thanks\relax}
\makeatother

\begin{document}
\fancyhead{}              

\maketitle

\begin{abstract}
Deep learning inference and training performance depends critically on GPU kernel efficiency. Modern compilers such as PyTorch Inductor automatically generate GPU kernels from high-level model code, but frequently underperform expert-written implementations by wide margins. Recent LLM-assisted optimizers explore these opportunities through kernel generation and refinement; yet translating isolated kernel improvements into model-level gains remains a challenge. We present KernelOPT, a profiling-guided multi-agent system for GPU kernel optimization. It inspects generated Triton kernels and external library calls in compiler artifacts to guide synthesis and refinement, then assembles and evaluates the resulting model. Four stages of validation combine execution checks, multi-seed numerical comparison, model-level checking with a float64 fallback and end-to-end performance gating ($\gamma{=}1.03$). The system supports PyTorch \texttt{nn.Module}s, standalone Triton kernels, and Helion kernels. It is evaluated on all 250 KernelBench problems on NVIDIA H200 which includes 100 single-operator problems (Level~1), 100 composite operators (Level~2), and 50 model architectures (Level~3). KernelOPT achieves geometric mean speedups over \texttt{torch.compile} of 1.40$\times$, 1.15$\times$ and 1.07$\times$ respectively across L1, L2 and L3. Among problems exceeding 1.03$\times$ speedup, the geomeans are 2.54$\times$ (36~L1), 1.84$\times$ (23~L2), and 1.37$\times$ (11~L3). These gains include arithmetic-precision changes and task-specific algebraic simplifications under the evaluated numerical tolerances. The evaluation also reveals discrepancies between isolated kernel improvements and end-to-end performance, motivating model-level measurement alongside kernel-level optimization.
\end{abstract}

\section{Introduction}

Efficient GPU kernels are the foundation of performant deep learning.
PyTorch Inductor~\citep{ansel2024pytorch}, the default
\texttt{torch.compile} backend, combines generated Triton~\citep{tillet2019triton}
kernels with external library calls. Compiler-generated code can still offer
opportunities for improvements in scheduling, fusion and algebraic simplification.
Expert implementations such as
FlashAttention~\citep{dao2022flashattention, dao2023flashattention2, shah2024flashattention3}
illustrate the value of specialized algorithms and GPU-aware implementation,
but developing such kernels requires substantial engineering effort.

Recent work has applied LLM agents to automate this process.
\textbf{KernelAgent}~\citep{kernelagent2026} translates
PyTorch models to Triton kernels via multi-agent code-to-code generation
with NCU-guided optimization.
\textbf{AccelOpt}~\citep{zhang2026accelopt} introduces an iterative
planner--executor--summarizer agent loop with optimization memory for
NKI kernels on AWS~Trainium.
\textbf{K-Search}~\citep{ksearch2026} formulates kernel generation as
planning over a co-evolving LLM world model with tree-structured search.
These systems differ in optimization scope, profiling feedback and verification.
KernelOPT studies their combination within a compiler-artifact workflow.

A compiled model contains both generated kernels and external library calls.
KernelOPT uses this structure to identify optimization targets and provide
context for synthesis. In the full-artifact replacement path, external calls
remain outside the targeted Triton bodies. This restriction trades search coverage,
including some cross-boundary fusion opportunities, for a narrower search space.
It is not evidence that library kernels are universally optimal. The tracing
path can force Triton GEMM templates before identifying remaining extern calls;
therefore dispatch preservation refers to calls remaining in that traced
artifact, not necessarily the dispatch chosen by an unconstrained compiler.
Standalone synthesis can also change dispatch using the traced operation groups.

We present an implementation combining compiler-artifact decomposition,
profiling-guided agent search and end-to-end checks; an evaluation on all
250 KernelBench problems; and an artifact-level analysis of the difference
between isolated and model-level measurements. Code is available at
\url{https://github.com/TorchedHat/KernelOPT}.

\section{Related Work}

\paragraph{Compiler-based and manual optimization.}
Halide~\citep{ragan2013halide} and TVM with Ansor~\citep{chen2018tvm,
zheng2020ansor} separate algorithm from schedule to enable large search
spaces. Triton uses block-level programming with automatic compiler
scheduling; PyTorch Inductor~\citep{ansel2024pytorch} integrates Triton
codegen into \texttt{torch.compile} but constrains autotuning to fixed
tile sizes and fusion patterns.
ThunderKittens~\citep{spector2024thunderkittens} provides tile-level
CUDA abstractions for hand-writing high-performance attention and SSM
kernels. Mirage~\citep{wu2025mirage} applies multi-level
superoptimization with probabilistic equivalence verification.

\paragraph{LLM-driven kernel optimization.}
KernelAgent~\citep{kernelagent2026} combines multi-agent generation,
NCU profiling, and beam search for Triton synthesis.

AccelOpt~\citep{zhang2026accelopt} introduces the iterative
planner--executor--summarizer architecture with beam search and
optimization memory for NKI kernels on AWS~Trainium; we adapt this
iterative framework to NVIDIA GPUs and extend it with Inductor-aware
synthesis, model-level verification, and guided fallback.
CudaForge~\citep{cudaforge2025} uses a two-agent Judge--Coder system.
K-Search~\citep{ksearch2026} uses LLM-guided tree search with a
co-evolving world model.
Astra~\citep{astra2025} employs a multi-agent architecture with
iterative profiling for CUDA kernel optimization but targets existing
production kernels rather than compiled-model synthesis.
OptiML~\citep{bhattacharjee2026optiml} combines
program synthesis with MCTS-guided refinement for CUDA kernels.
AutoKernel~\citep{jaber2026autokernel} applies an iterative keep/revert
loop with model-level profiling and a five-stage correctness harness.
KernelOPT combines compiler-artifact processing, NCU-guided planning and
model-level checks within this line of work.
\paragraph{Benchmarks.}
KernelBench~\citep{ouyang2025kernelbench} provides 250~problems across
three difficulty levels with standardized evaluation using
\texttt{torch.allclose(rtol=$10^{-4}$, atol=$10^{-4}$)}. We evaluate
on all 250~problems plus 20~tutorial kernels.

\section{Problem Formulation}

\begin{definition}[Kernel Optimization]
\label{def:kernel_opt}
Let $M_0 = \langle \{k_1, \dots, k_m\}, L \rangle$ denote a compiled
model produced by \texttt{torch.compile}, where $\{k_i\}$ are
Triton-generated sub-kernels and $L$ are extern library calls
(cuBLAS, cuDNN). During an individual Triton body replacement, we hold
$L$ fixed. For one optimization target,
let $k_0$ be its baseline standalone implementation and let $\calK$ be the
space of syntactically valid Triton programs. Let $\tau(k)$ denote isolated
latency, estimated from callable timing or NCU duration as described in
Section~\ref{sec:beam-notes}. Let $M(k_i \!\to\! k'_i)$ denote the
re-stitching operator that substitutes $k'_i$ into the model. For the
model context in which $k_0$ is replaced by $k$, let $\calV(k,k_0)$
denote the verification function. The ideal isolated objective is
\begin{equation}
    k^* = \arg\min_{k \in \calK} \; \tau(k) \quad \text{subject to} \quad \calV(k, k_0) = 1
    \label{eq:opt}
\end{equation}
\end{definition}

The search space $\calK$ is intractable to enumerate. We access it
through an LLM-based \emph{transformation operator}
$\calA_\theta: \calK \times \calP \to \calK$, where $\calP$ is the set of
optimization plans. The operator is parameterized by an
LLM~$\theta$ and takes a kernel $k$ and an optimization plan
$p \in \calP$ (derived from NCU profiling) and produces a candidate
$k' = \calA_\theta(k, p)$.

\begin{definition}[Four-Gate Verification Cascade]
\label{def:verification}
The intended cascade $\calV(k,k_0)$ combines four checks. The notation
$V_{\text{model}}(k,k_0)$ and $V_{\text{perf}}(k,k_0)$ refers to checks
on the model containing the replacement:
\begin{equation}
\begin{aligned}
\mathcal{V}(k, k_0)
&= \underbrace{V_{\text{stat}}(k) \;\wedge\; V_{\text{corr}}(k, k_0)}
   _{\text{per-candidate (Gates 1--2)}}
\;\wedge\;
\underbrace{V_{\text{model}}(k, k_0) \;\wedge\; V_{\text{perf}}(k, k_0)}
   _{\text{end-to-end (Gates 3--4)}}
\end{aligned}
\label{eq:verification}
\end{equation}
\end{definition}
The checks applied to a returned artifact depend on the execution path.
Search ranks candidates using isolated measurements, and failed
re-stitching checks can trigger targeted retries. The final verifier
can fall back to a weaker sanity check if corresponding inputs cannot
be reconstructed or if the candidate fails to execute on those inputs.
Section~\ref{sec:correctness} describes these paths.

\section{Method}

\subsection{System Overview}

KernelOPT supports standalone Triton or Helion~\citep{helion2025} inputs
and PyTorch model inputs. Both use Nsight Compute~\citep{nvidiansight}
profiling, strategy analysis, and an iterative plan--execute--check--profile
loop. The default PyTorch multi-kernel workflow has two routes.
\emph{Standalone synthesis} generates implementations from traced operation
groups, validates them against the original eager model, and optimizes
passing implementations in isolation before assembly. If no implementation
passes synthesis validation, the \emph{full-artifact route} extracts Inductor
output, tries applicable fusions, and optimizes eligible Triton bodies.
The model-level checks on proposed swaps can trigger targeted retries. When either
route produces a candidate model, the CLI attempts final numerical
verification against the original eager model. If verification
passes, it attempts end-to-end timing of the candidate, the eager model,
and the \texttt{torch.compile} baseline. If Inductor extraction yields no
Triton kernels, the CLI writes a \texttt{torch.compile} fallback and exits
before final benchmarking.

\begin{figure}[t]
  \centering
  \includegraphics[width=\textwidth]{Figures/model_final.pdf}%
  \caption{KernelOPT's PyTorch workflow. Tracing identifies generated
  Triton kernels and external library calls. Profiling guides isolated
  search. The model reaches end-to-end evaluation through standalone
  assembly or replacement within an Inductor artifact.}
  \label{fig:architecture}
\end{figure}

\paragraph{Inductor-aware synthesis.}
Tracing enables max-autotune and selects Triton GEMM and convolution
backends where those settings are available. It then identifies
\texttt{async\_compile.triton} blocks and external calls in the generated
source. The grouping heuristic merges up to six groups into one. For
larger inputs, it keeps template and extern groups separate while merging
other consecutive groups. The generator can rewrite the operations in a
group, including their dispatch and arithmetic.

For each resulting group, the system makes up to $K$ code-generation
attempts. It checks each generated implementation against the original
eager model using test inputs generated with PyTorch random seeds 0 and 42
and \texttt{torch.allclose} with
$\mathrm{rtol}=\mathrm{atol}=10^{-4}$. Passing implementations become
isolated optimization targets. The evaluation measures the combined
effect of synthesis and subsequent search.

\subsection{Agent Architecture}

KernelOPT adapts AccelOpt's planner--executor--summarizer
architecture~\citep{zhang2026accelopt} in a LangGraph state
machine~\citep{langgraph2024}. The \textbf{profiler} selects launch
settings and extracts throughput, duration, launch metrics, and
optimization rules from an Nsight Compute report (\texttt{ncu --set full}).
Strategy analysis assigns a heuristic tier from memory and compute
Speed-of-Light (SOL) throughput. It recommends parameter tuning if either
value exceeds 80\% and compares the two otherwise. For eligible
multi-kernel inputs, an LLM proposes a higher-level strategy.
The \textbf{planner} uses profiling evidence and prior directions to
generate plans. Its per-invocation hints can target different NCU rules,
and it deduplicates returned plans by direction within each batch.
The \textbf{executor} implements plans with error-guided retries.
The \textbf{summarizer} records experience from profiled candidates.

\paragraph{Executor retry budget.}
\label{sec:executor-budget}
For each plan, the pipeline can invoke the executor up to $K$ times,
including the first invocation. Within an invocation, the executor's
submission counter has a limit of $K$. Failed static checks return
feedback within the same conversation. The counter advances once per
agent turn containing a submission, so this limit counts submission
rounds. If an invocation returns no usable implementation or the
standalone dry run fails, the next invocation receives the error as
feedback. An implementation that passes the dry run proceeds to isolated
numerical checking. If that check fails and $K>1$, the pipeline allows
one additional executor invocation with a submission-round limit of one
and proceeds directly to numerical checking if the repaired submission
passes static validation. The main evaluation sets $K=4$.
Standalone synthesis uses $K$ as the code-generation attempt limit per
traced group. Model-level
re-stitching retries have their own limit (see \S\ref{sec:correctness}).

\subsection{Profiling-Guided Beam Search}

For one isolated target, the search starts at the baseline kernel $k_0$.
We represent the explored variants as nodes of a tree rooted at $k_0$.
An edge $(u,v)$ represents a proposed transformation
$k_v=\mathcal A_\theta(k_u,p)$ induced by plan $p$. An active entry carries
a kernel, its profiling context $\pi$, and a chain identifier $g$.
The beam determines which variants to expand. The persisted tree records
available candidate ancestry. A separate best-kernel
record determines the isolated result returned by the search.

\begin{algorithm}[t]
\caption{Profiling-guided beam search for one isolated target ($B>1$)}
\label{alg:beam}
\begin{algorithmic}[1]
\REQUIRE Baseline $k_0$, beam width $B$, iteration limit $T$, nominal plan budget $N$,
executor parameter $K$, initial memory $\calM$
\ENSURE Isolated search result $\hat{k}$
\STATE $\pi_0 \gets \mathrm{Profile}(k_0)$
\STATE $\calB_0 \gets \{(k_0,\pi_0,g_0)\}$,\quad $\hat{k} \gets k_0$
\FOR{$t=1,\ldots,T$}
    \STATE $n_t \gets \max(1,\lfloor N/|\calB_{t-1}|\rfloor)$
    \STATE $P_t \gets \mathrm{Plan}(\calB_{t-1},\calM,n_t)$
    \STATE Append a tuning plan if $P_t\neq\emptyset$ and all plans are complex
    \STATE $\calC_t \gets \emptyset$
    \FOR{each $(k,p,g)\in P_t$}
        \STATE $k' \gets \mathrm{ExecuteAndCheck}(k,p,k_0,K)$
        \IF{$k'\neq\bot$}
            \STATE $\pi' \gets \mathrm{Profile}(k')$
            \STATE $\calC_t \gets \calC_t\cup\{(k',\pi',g)\}$
        \ENDIF
    \ENDFOR
    \STATE Update $\calM$ from profiling outcomes and implementation failures
    \STATE Update the incumbent and global record $\hat{k}$ using isolated timings
    \STATE $\calB_t \gets \mathrm{DiverseSelect}(\calC_t,B)$ using Eq.~\ref{eq:diverse}
    \STATE Apply the active-set fallback if $\calB_t$ is empty
    \IF{two non-improving rounds in a row or a nonempty round of only low-confidence candidates}
        \STATE \textbf{break}
    \ENDIF
\ENDFOR
\STATE \textbf{return} $\hat{k}$
\end{algorithmic}
\end{algorithm}

\paragraph{Planning and candidate screening.}
Algorithm~\ref{alg:beam} summarizes the multi-chain search. The planner
requests $n_t$ plans per active entry, using its source, profiling context,
and prior directions. Thus $|\calB_{t-1}|n_t$ is the number of planner
requests before failures or deduplication. The code may append one tuning
plan if the nonempty plan list contains only complex plans. With $N=B=4$,
there are at most four planner requests per iteration for each isolated
target, plus any appended tuning plan. \textsc{ExecuteAndCheck} applies
the validation and retry procedure in \S\ref{sec:executor-budget}.
It returns $\bot$ if that procedure yields no candidate that passes
isolated numerical checking. Only passing candidates proceed to profiling.

\paragraph{Beam scoring and selection.}
Let $\calC_t^{\mathrm{ok}}$ contain candidates with a passing isolated
correctness flag and finite profiling latency. The beam uses a UCB-style
score when the search tree is available. Let $s(c)$ denote the recorded
speedup of candidate $c$ relative to the isolated baseline, $e(c)$ its
recorded expansion count (zero if no matching node is found), and
$E_t$ the sum of expansion counts over correct tree nodes. For the profiled
runs, $\tau_{\rm ncu}(c)$ denotes the candidate's NCU duration. The
ranking score is
\begin{equation}
q_t(c)=
\begin{cases}
\max(s(c),0)+1.4\sqrt{\dfrac{\log(1+E_t)}{1+e(c)}},
    & \text{if the tree is available},\\[4pt]
-\tau_{\rm ncu}(c), & \text{otherwise}.
\end{cases}
\label{eq:beam-score}
\end{equation}
For each chain identifier $g$, let $r_g$ be the first candidate from that
chain in descending score order. An unassigned chain identifier is grouped
under the default chain in this pass. Let
$R_t=\{r_g: g\text{ is represented in }\calC_t^{\mathrm{ok}}\}$.
With $\operatorname{Top}_j(S;q)$ returning up to $j$ highest-scoring
elements of $S$, selection takes representatives first and then fills
unused slots:
\begin{equation}
\begin{aligned}
D_t &= \operatorname{Top}_B(R_t;q_t),\\
\mathrm{DiverseSelect}(\calC_t,B)
    &= D_t \cup \operatorname{Top}_{B-|D_t|}
       (\calC_t^{\mathrm{ok}}\setminus D_t;q_t).
\end{aligned}
\label{eq:diverse}
\end{equation}
The ordered selection becomes the next active set. If candidates exist but
none is viable, the graph retains the previous live chains. If the candidate
set is empty, the graph uses its single-active-entry update.

\paragraph{Implementation notes on final selection.}
\label{sec:beam-notes}
The incumbent update compares isolated callable timings when every
candidate in the round has such a measurement. Otherwise it compares NCU
durations. The graph separately tracks a global best using the same choice
of metric for the round and returns that record when available. The metric
can change between rounds. The search result therefore comes from these
state updates, rather than a minimum over the final beam or a ranking by
full-model runtime.

An optional in-search re-stitching check records model-level correctness
without changing the isolated correctness flag. Outside the isolated
search, the full-artifact route checks a proposed swap before retention
and can run a new targeted search after failure. The standalone-synthesis
route assembles its selected outputs before the final model check.
Section~\ref{sec:correctness} describes both routes' model-level checks.

\subsection{Verification Cascade}
\label{sec:correctness}

\paragraph{Gates 1 \& 2: Static Validation and Multi-Seed Correctness.}
Dry-run execution catches syntax errors and runtime failures. The isolated
numerical checker uses seeds 0, 42 and 1337. When a standalone reference path
is supplied, it compares against that implementation at
\texttt{rtol=atol=$10^{-3}$}. This reference may itself be synthesized Triton;
it is not always the eager PyTorch model. Without such a reference, a
correctness agent generates a CPU reference and test specification, including
tolerances; tolerances are floored at $10^{-3}$ for outputs larger than
100,000 elements. Candidates that fail after the retry procedure in
\S\ref{sec:executor-budget} are discarded before profiling. The separate
synthesis validation uses the original eager model at $10^{-4}$.
These are finite tests, not equivalence proofs.

\paragraph{Gate 3: Model-level correctness ($V_{\text{model}}$).}
\label{sec:weight}
For corresponding inputs, the final verifier first attempts
\texttt{allclose(rtol=atol=$10^{-4}$)}. If that fails, it evaluates a double
precision reference. Let $d_{\rm ref}=\|r_{32}-r_{64}\|_\infty$,
$d_{\rm opt}=\|o_{32}-r_{64}\|_\infty$, and $a=\|r_{32}\|_\infty$.
For $d_{\rm ref}\geq10^{-8}$, the implemented acceptance condition is
\begin{equation}
\rho=\frac{d_{\rm opt}}{d_{\rm ref}},\qquad
(\rho\leq10)\ \lor\ (d_{\rm opt}\leq\max(5\cdot10^{-3},10^{-3}a)).
\label{eq:rho}
\end{equation}
When $d_{\rm ref}<10^{-8}$, dividing by the eager model's error
can make the ratio large even when the candidate's error is small.
The verifier instead scales the candidate's error by the magnitude
of the eager output and requires
$d_{\rm opt}/\max(a,10^{-12})\leq10^{-4}$.
These bounds can accept candidates that fail the initial element-wise
check.

If the verifier cannot reconstruct the reference inputs or the candidate
fails on them, it can run the candidate on inputs from its own generator.
In that case, it checks output count, shape, and dtype against the
reference and screens the values for nonfinite or nearly constant outputs.

When the candidate exposes model parameters as arguments, the verifier
attempts to recover them using a capture backend. Appendix~\ref{app:verification}
describes how it matches data inputs to captured arguments.

\paragraph{Precision instructions.}
The planner prompt warns against TF32 permission and BF16 casts on GEMM
operands. However, the runtime executor context for detected matmul templates
that are not classified as reductions explicitly recommends
\texttt{allow\_tf32=True}. These instructions conflict and are not an enforced
arithmetic-precision policy. Numerical acceptance, rather than a uniform
precision restriction, governs candidate retention. The precision-changing
results below must be interpreted accordingly; the prompts do not establish
that candidates preserve the baseline arithmetic.

\paragraph{Gate 4: Performance gate ($V_{\text{perf}}$).}
For a proposed swap in the full-artifact route, let $M_{\rm cur}$ be the
current assembled model and $M_{\rm swap}$ the model containing the
replacement. The gate uses subprocess-isolated CUDA-event timings
$\tau_M$ from \texttt{do\_bench} (25\,ms warmup, 100\,ms repetition):
\begin{equation}
    V_{\text{perf}}(M_{\rm swap},M_{\rm cur}) =
    \mathbb{1}\!\left[\tau_M(M_{\rm swap}) \leq
    \gamma \cdot \tau_M(M_{\rm cur})\right],\quad \gamma=1.03.
    \label{eq:perf}
\end{equation}
The gate allows up to a 3\% increase in runtime for a swap. The current
model can already include earlier replacements. A failed check can
trigger a new targeted search, with three retries in the benchmark
runner and two by default. The per-swap helpers can allow the search to
continue when they catch an internal error. The CLI separately attempts
a final model benchmark against \texttt{torch.compile} and falls back
when a measured speedup is below $0.97\times$.

\subsection{Experience Memory}

Following AccelOpt~\citep{zhang2026accelopt}, KernelOPT maintains a
bounded FIFO queue of capacity $Q$ storing experience items curated by
asymmetric speedup thresholds. For positive finite speedup $s(e)$,
$s(e)=\tau_{\rm baseline}/\tau_{\rm candidate}$; $s(e)^{-1}$ is the slowdown factor:
\begin{equation}
    \text{store}(e) = \mathbb{1}[s(e) \geq s_+] \;\vee\; \mathbb{1}[s(e)^{-1} \geq s_-]\,,
    \label{eq:memory}
\end{equation}
where $s(e)$ is the speedup and $s_+ = 1.05$, $s_- = 1.20$. The
asymmetry reflects that the thresholds select at least a 5\% speedup or a 1.20 runtime
slowdown factor for memory admission; they are not significance tests.
Execution/correctness failure records can bypass this speedup-based curation.

The implementation also includes a \emph{cross-run strategy tracker} that
flags strategies with $<$30\% success after $\geq$3 attempts as AVOID, and
ships \emph{benchmark-derived guidelines} as planner context. These are
suggestions rather than enforced constraints. Their individual effects are
not isolated in the evaluation; reuse on new targets requires accounting for
the provenance of this context.

\section{Experimental Setup}
\label{sec:setup}

Each problem uses a bare NVIDIA H200 GPU (132 SMs), with multiple problems
scheduled on different GPUs where available. The search backend is Claude
Sonnet~4.6 through Google Cloud Vertex AI.

\paragraph{Baseline selection.}
\label{sec:baselines}
For the primary baseline, we compile the original model with
\texttt{torch.compile} and enable Inductor's \texttt{max\_autotune}.
We warm up the compiled model before measuring its runtime. We also
report speedups over PyTorch eager execution.

We evaluate on the full KernelBench
benchmark~\citep{ouyang2025kernelbench}: Level~1 (100~single-op
kernels), Level~2 (100~composite operators), and Level~3
(50~model architectures), plus 10~Triton and 10~Helion tutorial
kernels (270~total problems). The model timing harness runs eager,
compiled, and candidate callables in separate subprocesses and draws
inputs from each callable's input generator. It measures runtime with
\texttt{do\_bench} (25\,ms warmup, 100\,ms repetition). The separate
model-level correctness check attempts to use corresponding inputs and
model parameters as described in Section~\ref{sec:correctness}.
The reported total compute cost is ${\sim}$1{,}100 GPU-hours on H200
(L1: ${\sim}$145\,h on 1~GPU; L2: ${\sim}$210\,h on 2~GPUs;
L3: ${\sim}$730\,h on up to 8~GPUs in parallel, with individual
models ranging from 6~min to 66~h).

The full-system search uses $T=5$, $N=4$, $K=4$, and $B=4$ per isolated
target, with early stopping and additional synthesis and model-level
retries. The profiler defaults to NCU kernel replay. The CLI benchmarks
the assembled model end to end after it passes the final correctness
check.

\section{Results and Discussion}

\subsection{Main Results}

Table~\ref{tab:results} summarizes one terminal outcome per problem.
A case is classified as optimized when the kept speedup
exceeds $1.03\times$, matched when it falls within the $[0.97,1.03]$
reporting band, and fallback-labelled when verification fails, performance
is rejected, or extraction yields no Triton kernels. For the geometric
means, we use the raw logged ratio when a matched outcome is displayed as
approximately $1.00\times$, and assign fallback-labelled cases $1\times$.

Three fallback-labelled L1 runs (041, 047, 049) retain generated Triton
\texttt{final\_model.py} files. Their aggregate factors reflect the
baseline-substitution convention. L3-018 and L3-044 each have more
than one terminal report. We use the last report for each. Which attempt
produced the saved artifacts is unclear for these two problems.

\begin{table*}[t]
\centering
\footnotesize
\setlength{\tabcolsep}{4pt}

\resizebox{\textwidth}{!}{%
\begin{tabular}{@{} lrrrrrr @{\hspace{2em}} lrrrr @{}}
\toprule
\multicolumn{7}{@{}l}{\textbf{A.\ Summary}} & \multicolumn{5}{@{\hspace{2em}}l}{\textbf{B.\ Speedup distribution (70 opt.)}} \\
\cmidrule(r){1-7} \cmidrule(l){8-12}
Level & Total & Opt. & Matched & Fall. & Geo. & Max & Range & L1 & L2 & L3 & Tot. \\
\midrule
L1  & 100 & 36 & 48 & 16 & \textbf{1.40}$\times$ & 88.63$\times$ & $\geq$10$\times$ & 4 & 1 & 0 & 5 \\
L2  & 100 & 23 & 43 & 34 & \textbf{1.15}$\times$ & 33.11$\times$ & $[5, 10)$        & 6 & 4 & 0 & 10 \\
L3  &  50 & 11 &  4 & 35 & \textbf{1.07}$\times$ &  4.12$\times$ & $[2, 5)$         & 7 & 0 & 2 & 9 \\
    &     &    &    &    &                       &               & $[1.5, 2)$       & 2 & 2 & 1 & 5 \\
    &     &    &    &    &                       &               & $[1.1, 1.5)$     & 8 & 8 & 4 & 20 \\
\cmidrule(r){1-7}
All & 250 & 70 & 95 & 85 & \textbf{1.23}$\times$ & 88.63$\times$ & $(1.03, 1.1)$    & 9 & 8 & 4 & 21 \\
\bottomrule
\end{tabular}%
}

\caption{Terminal log outcomes and baseline-substituted speedups. Panel A
assigns fallback-labelled cases $1\times$; Panel B uses the printed factors
of the 70 optimized outcomes. Rounded factors limit boundary precision.
These results are not a precision-matched evaluation or a measurement of every
saved final-model artifact.}
\label{tab:results}
\end{table*}

\paragraph{Speedup analysis.}
Table~\ref{tab:results} shows that a few workloads account for the
largest speedups. Table~\ref{tab:traces} compares isolated and model-level
measurements.

\emph{Structured workloads.} L1-012 replaces diagonal matrix
multiplication with row scaling (88.63$\times$). L2-018 simplifies a
singleton-reduction chain after a linear layer (33.11$\times$).
The saved kernels for lower-triangular multiplication (L1-015,
20.26$\times$) and upper-triangular multiplication (L1-014, 14.04$\times$)
also cast matrix operands to FP16. Their timings include both structural
and precision changes. L1-085 uses a specialized Triton kernel for an
asymmetric depthwise 2D convolution (11.38$\times$).

\emph{Fusion with precision changes.} L2-064 (8.30$\times$) and L2-022
(7.18$\times$) contain matrix products with fused epilogues. Both explicitly
permit TF32 in \texttt{tl.dot}; the results do not isolate the benefit of fusion
from arithmetic precision and other implementation changes.

Appendix~\ref{app:tutorials} reports results for standalone tutorial kernels.

\paragraph{Precision and scope limitations.}
Among the 70 optimized outcomes, 16 matrix-product tasks explicitly use
FP16/BF16 operands or permit TF32 in their generated source. A seventeenth,
L3-002, omits the \texttt{tl.dot} input-precision setting, leaving it to Triton's
default and environment configuration. Appendix~\ref{app:precision} lists these
17 cases. Assigning just these cases $1\times$ while retaining all denominators
reduces the L1--L3 geomeans to 1.14$\times$, 1.07$\times$ and 1.04$\times$;
they account for 57.4\% of the summed log-speedup over all 250 problems.
This sensitivity analysis is not a matched-precision re-timing, an exhaustive
classification of executed arithmetic, or a measurement of the remaining gain
attributable to dispatch awareness. Passing numerical tolerances does not
establish equal arithmetic precision.
The L2-018 wrapper recomputes the weight and bias reductions inside the
timed call. Its arithmetic reduction depends on the singleton reductions
in this benchmark task (Appendix~\ref{app:examples}).

\paragraph{Robustness of logged aggregates.}
Baseline-substituted median speedups are 1.01$\times$, 1.00$\times$, and
1.00$\times$ for L1--L3. Excluding factors at least $10\times$ leaves 96, 99
and 50 problems, with geomeans 1.245$\times$, 1.112$\times$, and 1.072$\times$.
This is a descriptive tail-exclusion analysis, not a matched-precision control.
The reported GPU-hours and token counts describe the cost of optimization.

\subsection{Speedup Baselines: vs.\ Compile and vs.\ Eager}

We report geometric mean speedups over \texttt{torch.compile} and
PyTorch eager execution, both across the benchmark and for optimized
outcomes only (Table~\ref{tab:speedups}). The eager comparison covers 238
problems because five runs stopped at final verification and seven took
an early compiler fallback before timing.

\begin{table}[h]
\centering
\footnotesize
\begin{tabular}{@{}lrrrr@{}}
\toprule
\textbf{Level} & \textbf{vs compile (all)} & \textbf{vs compile (opt-only)} & \textbf{vs eager (all timed)} & \textbf{vs eager (opt-only)} \\
\midrule
L1 (100) & 1.40$\times$ & 2.54$\times$ & 1.72$\times$ & 2.64$\times$ \\
L2 (100) & 1.15$\times$ & 1.84$\times$ & 1.79$\times$ & 2.82$\times$ \\
L3 (50)  & 1.07$\times$ & 1.37$\times$ & 1.56$\times$ & 1.69$\times$ \\
\bottomrule
\end{tabular}
\caption{Geometric mean speedups over \texttt{torch.compile} and eager
PyTorch. The all-problem compile comparison counts fallback outcomes
at $1\times$. For timed fallback cases, the eager comparison uses the
measured compile speedup over eager. The all-timed eager column covers
99 L1, 100 L2, and 39 L3 problems. The optimized-only columns cover
36 L1, 23 L2, and 11 L3 problems.}
\label{tab:speedups}
\end{table}

Across problems with timings for both baselines, \texttt{torch.compile}
itself achieves geometric mean speedups over eager of $1.22\times$,
$1.56\times$, and $1.43\times$ on Levels~1--3, respectively.

\subsection{Fallback Analysis}

The terminal logs classify 73 performance-regression fallbacks, five final
verification failures and seven no-generated-kernel cases. The 73 performance
cases comprise 15 L1, 34 L2 and 24 L3 problems. Name-based GEMM/Conv labels
are not measurements of library runtime dominance or explanations of failure,
so we do not use them as causal categories. L1-049 records 2.466 ms for compile
and 2.668 ms for the candidate (0.92$\times$); its saved model still contains
Triton code. This demonstrates the need to distinguish a rejected logged
candidate from the saved artifact and to qualify the baseline-substitution metric.

\subsection{The Multi-Kernel Composition Gap}
\label{sec:composition-gap}

The full-artifact route reports per-kernel optimization summaries for
126 of the 250 problems. In 37 of these problems (29\%), at least one
reported isolated speedup exceeds $1.03\times$, while the final outcome
is either matched (21 cases) or a performance fallback (16 cases).
The summaries list optimization targets. They can reset
rejected swaps to $1\times$ or retain an isolated speedup after a
stitching exception. For example, L2-034 reports $30.75\times$ for its
isolated kernel and $0.994\times$ end to end.

Dispatch overhead, layout conversions, and launch configuration can
affect the relationship between the two measurements. The saved timings
do not quantify these effects.

For a single replacement in an otherwise unchanged serial execution,
Amdahl's prediction is $S=1/((1-f)+f/s)$, where $f$ is the replaced kernel's
fraction of the original model runtime and $s$ its speedup under the same
inputs and measurement context. Neither kernel counts nor isolated NCU
latencies against a synthesized standalone baseline provide that $f$.
Here the isolated and E2E comparisons can also have different baselines;
therefore subtracting this prediction from an observed E2E factor would not
identify stitching loss without additional in-model measurements. On the
Inductor re-stitching path the code strips the autotune decorator and uses
the first listed config to adjust a size hint; this does not preserve a
measured winning config. Its contribution to any particular lost gain is
unmeasured. The standalone assembly route has a separate source-selection
limitation: optimization writes directories using 30-character kernel names,
whereas assembly looks them up using 60 characters and falls back to initial
source if no match is found. This can also make assembled and isolated code
differ; its incidence and timing effect have not been measured.

\subsection{Level 3 Outcomes and Ablation Study}

Level 3 contains 50 model architectures; eleven have optimized terminal outcomes
and the baseline-substituted geomean is $1.07\times$. The experiments do not
measure the fraction of baseline execution time attributable to optimizable
sub-kernels, so architecture names alone cannot explain these outcomes.

The ablation compares the full system with five flag variants on a
shared random sample of 50 problems. Each variant contributes one run
per problem. Appendix~\ref{app:ablation} gives the sampling procedure
and explains what the flags change.

\section{Conclusion}
\label{sec:conclusion}

KernelOPT uses compiler output to identify where to search and measures
the resulting implementation in the context of the full model. The two
routes expose different opportunities. Synthesis can exploit structure in
the original computation and avoid substantial work, as in the diagonal
matrix product. The full-artifact route can improve individual Triton
kernels within an existing implementation. In several cases, gains measured
in isolation coincided with little or no improvement in model latency.
Isolated timings guide the search, while end-to-end timings show whether
the assembled model runs faster.

\paragraph{Limitations.}
The evaluation covers fixed-shape workloads on one GPU type with one LLM
backend. We did not assess overlap between the public benchmark and the
LLM's training data. Several of the largest gains combine algorithmic
rewrites or scheduling changes with reduced arithmetic precision.
The experiments therefore leave open how much speedup remains at matched
precision and how much comes from synthesis or subsequent search. With one
reported outcome per problem, the study leaves run-to-run variability and
performance relative to other agents at equal cost unmeasured.

\paragraph{Future work.}
Measuring the synthesized model before search and the assembled model after
search under a common precision policy would help separate their
contributions. Repeating search and independently timing its winners on
shared inputs would show how stable the gains are. Comparisons with
simpler search methods and other agents under matched GPU and token
budgets would test the value of profiling-guided search. In-model profiling
of proposed swaps could also reveal when a local speedup is large enough to
affect model latency.

\subsection*{Acknowledgments}

We thank the Red Hat PyTorch Engineering team for their support and for
providing the compute resources used in this work. We are grateful to
Joseph Groenenboom and Ali Raza for their project direction and technical
guidance throughout the internship, and to Arkadip Maitra for reviewing
early drafts.

\subsection*{AI Use Statement}

The authors used LLM assistants to edit prose, find related work, and
draft portions of the manuscript. The authors checked references against
primary sources, reviewed and revised the AI-assisted text, and take
responsibility for the paper and its artifacts. KernelOPT's use of an LLM
as part of the optimization method is described in Section~\ref{sec:setup}.

\subsection*{Ethics Statement}

This work concerns performance optimization of GPU kernels and does not
involve human subjects, personally identifiable information, or
sensitive data. The KernelBench benchmark used for evaluation is a
publicly available, permissively licensed research artifact. Potential
societal considerations are limited to the compute and energy footprint
of the optimization procedure: our full evaluation consumed
${\sim}1{,}100$ H200 GPU-hours and ${\sim}115$M LLM tokens. We note that
the goal of the system is producing faster kernels hence reducing the
inference-time energy cost of downstream models, and that the verification checks are intended to detect failures, with the limitations
described in the verification section; they do not prove equivalence. We declare no conflicts of interest.

\subsection*{Reproducibility Statement}

Algorithm~\ref{alg:beam} describes the isolated search.
Section~\ref{sec:setup} gives the search settings and timing protocol.
The full-system settings include $T{=}5$, $N{=}4$, $K{=}4$, $B{=}4$,
$\gamma{=}1.03$, $Q{=}8$, $s_+{=}1.05$, and $s_-{=}1.20$. The
float64 check uses a ratio threshold of 10 alongside the additional
error bounds in Section~\ref{sec:correctness}.
The evaluation uses public KernelBench problems with the custom numerical
acceptance rules described above. Appendix~\ref{app:ablation} lists the
shared 50-problem ablation sample.
The complete codebase is available at
\url{https://anonymous.4open.science/r/KernelOPT-93A9}.

\begin{small}
\bibliography{references}

@misc{nvidiansight,
  title={{NVIDIA Nsight Compute}},
  author={{NVIDIA Corporation}},
  howpublished={\url{https://developer.nvidia.com/nsight-compute}},
  note={Performance Profiler for CUDA Applications},
  year={2024}
}

@misc{langgraph2024,
  title={{LangGraph}: Building Language Agents as Graphs},
  author={{LangChain Inc.}},
  howpublished={\url{https://github.com/langchain-ai/langgraph}},
  year={2024}
}

@inproceedings{tillet2019triton,
  title={{Triton}: An Intermediate Language and Compiler for Tiled Neural Network Computations},
  author={Tillet, Philippe and Kung, H. T. and Cox, David},
  booktitle={Proceedings of the 3rd ACM SIGPLAN International Workshop on Machine Learning and Programming Languages (MAPL)},
  pages={10--19},
  year={2019}
}

@inproceedings{ansel2024pytorch,
  title={{PyTorch 2}: Faster Machine Learning Through Dynamic {Python} Bytecode Transformation and Graph Compilation},
  author={Ansel, Jason and Yang, Edward and He, Horace and Gimelshein, Natalia and Jain, Animesh and Voznesensky, Michael and Bao, Bin and Bell, Peter and Berber, David and Burber, Matthias and others},
  booktitle={Proceedings of the 29th ACM International Conference on Architectural Support for Programming Languages and Operating Systems (ASPLOS)},
  year={2024}
}

@misc{helion2025,
  title={{Helion}: A {DSL} for Low-Level {Triton} Kernel Programming},
  author={Ansel, Jason and others},
  howpublished={\url{https://github.com/pytorch-labs/helion}},
  note={PyTorch Labs},
  year={2025}
}

@inproceedings{dao2022flashattention,
  title={{FlashAttention}: Fast and Memory-Efficient Exact Attention with {IO}-Awareness},
  author={Dao, Tri and Fu, Daniel Y. and Ermon, Stefano and Rudra, Atri and R{\'e}, Christopher},
  booktitle={Advances in Neural Information Processing Systems (NeurIPS)},
  volume={35},
  pages={16344--16359},
  year={2022}
}

@inproceedings{dao2023flashattention2,
  title={{FlashAttention-2}: Faster Attention with Better Parallelism and Work Partitioning},
  author={Dao, Tri},
  booktitle={International Conference on Learning Representations (ICLR)},
  year={2024}
}

@misc{shah2024flashattention3,
  title={{FlashAttention-3}: Fast and Accurate Attention with Asynchrony and Low-Precision},
  author={Shah, Jay and Bikshandi, Ganesh and Zhang, Ying and Thakkar, Vijay and Ramani, Pradeep and Dao, Tri},
  howpublished={arXiv preprint arXiv:2407.08691},
  eprint={2407.08691},
  archivePrefix={arXiv},
  year={2024}
}

@inproceedings{chen2018tvm,
  title={{TVM}: An Automated End-to-End Optimizing Compiler for Deep Learning},
  author={Chen, Tianqi and Moreau, Thierry and Jiang, Ziheng and Zheng, Lianmin and Yan, Eddie and Shen, Haichen and Cowan, Meghan and Wang, Leyuan and Hu, Yuwei and Ceze, Luis and Guestrin, Carlos and Krishnamurthy, Arvind},
  booktitle={13th USENIX Symposium on Operating Systems Design and Implementation (OSDI)},
  pages={578--594},
  year={2018}
}

@inproceedings{zheng2020ansor,
  title={{Ansor}: Generating High-Performance Tensor Programs for Deep Learning},
  author={Zheng, Lianmin and Jia, Chengfan and Sun, Minmin and Zhao, Zao and Yu, Cody Hao and Haj-Ali, Ameer and Wang, Yida and Yang, Jun and Zhuo, Danyang and Sen, Koushik and Gonzalez, Joseph E. and Stoica, Ion},
  booktitle={14th USENIX Symposium on Operating Systems Design and Implementation (OSDI)},
  pages={863--879},
  year={2020}
}

@inproceedings{ragan2013halide,
  title={{Halide}: A Language and Compiler for Optimizing Parallelism, Locality, and Recomputation in Image Processing Pipelines},
  author={Ragan-Kelley, Jonathan and Barnes, Connelly and Adams, Andrew and Paris, Sylvain and Durand, Fr{\'e}do and Amarasinghe, Saman},
  booktitle={Proceedings of the 34th ACM SIGPLAN Conference on Programming Language Design and Implementation (PLDI)},
  pages={519--530},
  year={2013}
}

@misc{kernelagent2026,
  title={{KernelFalcon}: Autonomous {GPU} Kernel Generation via Deep Agents},
  author={Wang, Laura and Cheng, Kaiming and Xu, Yilun and Pan, Jiafei and Zhou, Guanglei and Li, Rui and Zhang, Zhisong and Zhu, Kevin},
  howpublished={PyTorch Blog},
  note={\url{https://pytorch.org/blog/kernelfalcon-autonomous-gpu-kernel-generation-via-deep-agents/}},
  year={2025}
}

@misc{zhang2026accelopt,
  title={{AccelOpt}: A Self-Improving {LLM} Agentic System for {AI} Accelerator Kernel Optimization},
  author={Zhang, Genghan and Zhu, Shaowei and Wei, Anjiang and Song, Zhenyu and Nie, Allen and Jia, Zhen and Vijaykumar, Nandita and Wang, Yida and Olukotun, Kunle},
  howpublished={OpenReview preprint},
  note={\url{https://openreview.net/forum?id=SBS4NJHYjZ}},
  year={2026}
}

@misc{astra2025,
  title={{Astra}: A Multi-Agent System for {GPU} Kernel Performance Optimization},
  author={Wei, Anjiang and others},
  howpublished={arXiv preprint arXiv:2509.07506},
  eprint={2509.07506},
  archivePrefix={arXiv},
  year={2025}
}

@misc{jaber2026autokernel,
  title={{AutoKernel}: Autonomous {GPU} Kernel Optimization via Iterative Agent-Driven Search},
  author={Jaber, Jaber and Jaber, Osama},
  howpublished={arXiv preprint arXiv:2603.21331},
  eprint={2603.21331},
  archivePrefix={arXiv},
  year={2026}
}

@misc{cudaforge2025,
  title={{CudaForge}: An Agent Framework with Hardware Feedback for {CUDA} Kernel Optimization},
  author={Zhang, Zijian and Wang, Ruibo and Li, Siyuan and Luo, Yilong and Hong, Mingyi and Ding, Chen},
  howpublished={arXiv preprint arXiv:2511.01884},
  eprint={2511.01884},
  archivePrefix={arXiv},
  year={2025}
}

@misc{ksearch2026,
  title={{K-Search}: {LLM} Kernel Generation via Co-Evolving Intrinsic World Model},
  author={Cao, Shiyi and Mao, Ziming and Gonzalez, Joseph E. and Stoica, Ion},
  howpublished={arXiv preprint arXiv:2602.19128},
  eprint={2602.19128},
  archivePrefix={arXiv},
  year={2026}
}

@misc{bhattacharjee2026optiml,
  title={{OptiML}: An End-to-End Framework for Program Synthesis and {CUDA} Kernel Optimization},
  author={Bhattacharjee, Arijit and Ping, Heng and Le, Son Vu and Bogdan, Paul and Ahmed, Nesreen K. and Jannesari, Ali},
  howpublished={arXiv preprint arXiv:2602.12305},
  eprint={2602.12305},
  archivePrefix={arXiv},
  year={2026}
}

@inproceedings{ouyang2025kernelbench,
  title={{KernelBench}: Can {LLM}s Write Efficient {GPU} Kernels?},
  author={Ouyang, Anne and Guo, Simon and Arora, Simran and Zhang, Alex L. and Hu, William and R{\'e}, Christopher and Mirhoseini, Azalia},
  booktitle={Proceedings of the 42nd International Conference on Machine Learning (ICML)},
  year={2025}
}

@misc{spector2024thunderkittens,
  title={{ThunderKittens}: Simple, Fast, and Adorable {AI} Kernels},
  author={Spector, Benjamin F. and Arora, Simran and Singhal, Aaryan and Fu, Daniel Y. and R{\'e}, Christopher},
  howpublished={arXiv preprint arXiv:2410.20399},
  eprint={2410.20399},
  archivePrefix={arXiv},
  year={2024}
}

@inproceedings{wu2025mirage,
  title={{Mirage}: A {Multi-Level} Superoptimizer for Tensor Programs},
  author={Wu, Mengdi and Cheng, Xinhao and Liu, Shengyu and Shi, Chunan and Ji, Jianan and Ao, Man Kit and Velliengiri, Praveen and Miao, Xupeng and Padon, Oded and Jia, Zhihao},
  booktitle={Proceedings of the 19th USENIX Symposium on Operating Systems Design and Implementation (OSDI)},
  pages={235--252},
  year={2025}
}
\bibliographystyle{iclr2027_conference}
\end{small}

\appendix

\section{Agent Prompt Templates}
\label{app:prompts}

The following listings document the planner, executor, summarizer, profiler,
synthesis and fusion prompts, with some lengthy runtime references omitted.
They do not enumerate every LLM call: strategy analysis and the optional
correctness agent also use LLMs. The five optimization roles described in the
main text are not five independent models or a fixed five-call budget.

\subsection{Planner Agent}
The planner proposes one optimization plan per invocation. Invocations
within a planner batch run concurrently. The outer loop processes active
chains sequentially.
\begin{lstlisting}[style=prompt]
You are the Planner Agent in a GPU kernel optimization pipeline.
You receive a profiling context from Nsight Compute (NCU) and the kernel
source code. Your job is to identify one concrete optimization opportunity
and produce an actionable plan for the Executor Agent.
WORKFLOW -- follow these steps in order:
1. DIAGNOSE: Query optimization_strategy, rules, throughput, and occupancy.
Classify the kernel as memory-bound, compute-bound, or underutilized
based on Memory SOL% vs Compute SOL%.
2. INSPECT: Read the kernel source. Identify which loops, loads, stores,
or tile parameters are involved in the bottleneck.
3. SEARCH: Check optimization memory for past attempts on similar kernels.
Avoid directions that previously failed or regressed.
4. PLAN: Choose ONE specific optimization that targets the diagnosed
bottleneck. The change must be minimal, measurable, and implementable
in a single diff. Prefer @triton.autotune for parameter exploration.
5. SUBMIT: Call submit_plan with the plan.
Use your tools:
query_profiling_context(aspect) -- inspect specific profiling sections
aspects: optimization_strategy, rules, occupancy, memory_workload,
throughput, scheduler, launch, gpu_specs, instruction_stats,
multi_kernel_summary
get_kernel_source()
-- read the kernel source code
search_memory(query)
-- check past optimization experience
OPTIMIZATION GUIDANCE:


The kernel file may contain multiple GPU kernels (cuBLAS GEMM, cuDNN conv,
Triton @jit kernels, runtime kernels). Focus your plan on the @triton.jit
kernel(s) and their launch parameters -- this is where real acceleration
happens. Library calls (extern_kernels.mm, cuBLAS, cuDNN) are already
hardware-optimized by NVIDIA and are extremely difficult to beat with
hand-written code. Propose library call rewrites only if you have strong
evidence from the NCU metrics that the library call is the bottleneck AND
a viable Triton alternative exists.
When you have gathered enough information, call submit_plan(plan) with
your plan. The plan must:
- Target a specific, measurable bottleneck in the @triton.jit kernel(s)
- Reference the exact construct in the kernel source to modify
- Be implementable in one focused diff by the Executor Agent
- Include enough detail in step.change and step.implementation_hints
that the Executor does not need to re-read profiling data
Optimization strategies to consider (in priority order):
1. Add @triton.autotune with multiple configurations to explore tile sizes,
num_warps, and num_stages automatically at runtime
2. Adjust tile sizes (XBLOCK, YBLOCK, num_warps, num_stages) for better
occupancy
3. Improve memory access patterns (coalescing, vectorized loads,
L2 compression)
4. Reduce register pressure or branch divergence
5. Change eviction policies and cache hints
6. Fuse adjacent @triton.jit pointwise kernels (NOT library call fusion)
7. Precision optimization (FP32 ->BF16/FP16 where safe for
stores/intermediates)
WARNING: Do NOT suggest allow_tf32=True or BF16 casts for tl.dot
operands in GEMM kernels. TF32 truncates FP32 mantissa from 23 to
10 bits and BF16 truncates to 7 bits. For large K (>512), accumulated
error exceeds atol=1e-3. Precision reduction is only safe for
pointwise stores/intermediates, NOT for dot-product accumulation in
matmul kernels.
8. Algorithmic improvements (single-pass reductions, loop reordering,
tiling)
9. Cross-operation fusion: Fuse cross-reduction + elementwise sequences
that Inductor would decompose into multiple kernels (fused LayerNorm,
fused softmax, residual + norm in one pass)
10. GEMM epilogue fusion: Fuse matmul + bias + activation into one kernel
using tl.dot.
11. Warp specialization: Assign different warp groups to different tasks.
12. Persistent kernel scheduling: For workloads with many small tiles,
use a persistent kernel where each SM processes multiple tiles in a
loop rather than launching one CTA per tile.
INDUCTOR ANALYSIS -- when diagnosing torch.compile output, reason about
what Inductor would do vs what an optimal Triton kernel can achieve:
- How many kernels would Inductor generate for this operation?
- Which operations would it fuse? Which would it keep separate?
- Where are the memory traffic bottlenecks between Inductor's kernels?
- Can the Triton kernel fuse operations that Inductor keeps separate?
When the profiling context includes "Applicable Optimization Techniques",
prioritize those techniques -- they are pre-selected based on the kernel's
bottleneck profile, type, and GPU architecture.
[Helion-specific planning guidance appended at runtime for Helion kernels.]
\end{lstlisting}

\subsection{Executor Agent}
The Executor implements the Planner's optimization plan by modifying the
Triton kernel source. It receives validation and runtime errors as feedback,
with the nested invocation/submission budgets and separate correctness repair
described in \S\ref{sec:executor-budget}. The
prompt below is condensed; the full Helion configuration reference and
Triton API appendix (${\sim}$200 additional lines) are appended at
runtime.
\begin{lstlisting}[style=prompt]
You are the Executor Agent in a GPU kernel optimization pipeline.
You receive an optimization plan and a kernel source file.
Your job is to implement exactly the change described in the plan.
WORKFLOW -- follow these steps in order:
1. ANALYZE: Read the kernel source and optimization plan. Identify the
EXACT lines that need to change. State them.
2. REASON: In 2-3 sentences explain WHY this change improves performance,
referencing the NCU bottleneck from the plan's evidence.
3. IMPLEMENT: Make the MINIMAL change described in the plan. Modify ONLY
the identified lines. Do not refactor, restructure, or rewrite from
scratch. Copy the original kernel, then apply surgical edits.
4. VERIFY before submitting -- check each of these:
- Function signature matches the original exactly (name, params, order)
- All variables are defined (no new undefined constants)
- All tl.* calls exist in the Triton API
- If @triton.autotune added, config keys match existing constexpr params
- Code is syntactically valid Python
5. SUBMIT: Call submit_kernel(kernel_source, change_summary).
CRITICAL RULES (violations will be auto-rejected):
*** MOST IMPORTANT RULE ***
1. The function signature MUST be IDENTICAL to the original:
- SAME function name, parameter names, order, and count
- Do NOT add, remove, or rename any params
2. Every variable must come from: function parameters, computed locally,
or Triton builtins. Do NOT introduce undefined constants.
3. Submit ONLY the @triton.jit function (with decorators and body).
4. If adding @triton.autotune configs, the config keys must match
EXISTING constexpr parameters.
5. Do NOT restructure tiling (2D to 1D or vice versa).
6. Do NOT replace Triton computation with PyTorch calls.
TRITON AUTOTUNING:
@triton.autotune(configs=[...], key=[...], reset_to_zero=[...])
Critical: MASKING -- When adding @triton.autotune with configs that
increase XBLOCK/RBLOCK beyond the original, EVERY tl.load and tl.store
MUST have a mask= argument. This is the #1 cause of kernel crashes.
NUMERICAL STABILITY:
- Always accumulate reductions and tl.dot in float32
- Do NOT set allow_tf32=True on tl.dot unless explicitly asked
- Cast to output dtype only on the final tl.store
If submit_kernel returns a validation error, diagnose and resubmit.
[Full Triton API reference and Helion Config reference appended at runtime.]
\end{lstlisting}

\subsection{Summarizer Agent}
The Summarizer converts attempt records into structured experience items
for the memory queue. It proposes explanations from code differences and
profiling metrics; these explanations are planning context, not controlled
causal evidence.
\begin{lstlisting}[style=prompt]
You are the Summarizer Agent. Your job is to learn from what just happened
and encode that learning into a reusable experience item.


You do not apply rules. You observe, reason about causation, and write a
generalized insight.
## Summarization process
### Step 1 -- Identify the causal change
Diff the slow and fast kernels. Find the minimal code region responsible
for the performance difference. Verify against the actual diff.
### Step 2 -- Attribute the effect to metrics
Compare profiling_context before and after. Which metrics changed, by
how much? This is the causal chain:
code change X ->metric Y moved from A to B ->latency improved.
Do not speculate. Only attribute effects that appear in profiling data.
### Step 3 -- Generalize the insight
Ask: if a different kernel had the same profiling signature, would this
change help? Write strategy_description at that level of generality.
Reference profiling signals (metric names, values, NCU rules) not
specific variable names.
### Step 4 -- Write the pseudocode
Extract the key structural change as framework-neutral pseudocode.
Strip boilerplate. Keep loop structure, tile parameters, and access
patterns. Label framework-specific syntax.
### Step 5 -- Deduplication check
Scan existing memory queue. If an item with the same direction exists:
- Higher speedup: replace the old one
- Lower speedup: skip
- Different manifestation: keep both (append _v2 suffix)
## Output format
Two JSON objects separated by a blank line:
1. experience_item: {item_id, iteration, speedup, rewrite_type,
framework, direction, profiling_signal, strategy_title,
strategy_description, slow_pseudocode, fast_pseudocode,
applicable_when, do_not_apply_when, framework_notes}
2. memory_update: {action: "append"|"replace"|"skip",
replace_item_id, reason}
## Negative rewrite guidance
For regressions, explain which metrics got worse and why.
The strategy_description must explain the structural anti-pattern.
\end{lstlisting}

\subsection{Profiler Agent}
The Profiler chooses an NCU profiling configuration (kernel
name filter, launch skip/count, replay mode) by analyzing the kernel
source for framework markers, tensor initialization patterns, and
benchmark loop structure.
\begin{lstlisting}[style=prompt]
You are the Profiler Agent in a GPU kernel optimization pipeline.
Your job: determine the optimal NCU profiling configuration for a
given kernel.
For files with multiple kernel launches (Inductor-generated,
multi-kernel Triton): set launch_count to null (profile ALL kernels)
so the pipeline can discover the bottleneck.
Tools available:
read_kernel_source()
-- read the full kernel file
submit_ncu_config(...)
-- submit your configuration


When analyzing the source:
1. Identify the framework from decorators:
- @triton.jit or @tl.jit ->Triton
- @helion.kernel or @hl.kernel ->Helion
2. Set kernel_name:
- Triton / Helion: set kernel_name = "" (empty string).
Triton JIT-compiles kernels and gives them mangled names
that do NOT match the Python function name.
3. Infer launch_skip from the benchmark harness:
- Count torch/numpy tensor init calls before the first kernel
call. Each torch.randn / torch.zeros on GPU = 1 kernel launch.
4. Infer launch_count from steady-state repetitions:
- Look for a timing/profiling loop. Default to 1 if no loop found.
5. Extract kernel_args -- any CLI flags the script requires.
6. Choose replay_mode:
- Triton / Helion: always use 'application'.
JIT-compiled kernels are not compatible with NCU kernel replay.
Call submit_ncu_config once. Do NOT produce a human-readable summary.
\end{lstlisting}

\subsection{Codegen (Synthesis) Agent}
Used during Inductor-aware synthesis to generate standalone Triton
kernels from ATen operation specifications. The LLM receives operation
names, tensor shapes, and dtypes, and must produce a self-contained file
with an autotuned kernel, a wrapper function, and an input generator. The
prompt below is condensed; full code examples (${\sim}$150 lines) are
omitted.
\begin{lstlisting}[style=prompt]
You are an expert GPU kernel engineer. You write high-performance Triton
kernels from PyTorch operation specifications.
Given a description of what computation to implement (ATen ops, tensor
shapes, dtypes), you produce a complete, runnable Python file containing:
1. A @triton.jit kernel (optionally with @triton.autotune)
2. A kernel_function(*inputs) wrapper
3. A get_inputs() function returning sample input tensors
RULES:
- Import only: torch, triton, triton.language as tl
- Use float32 accumulators for numerical stability
- Include boundary masks for ALL tl.load and tl.store calls
- Include @triton.autotune with at least 4 configs
- CRITICAL: When using @triton.autotune with variable BLOCK sizes,
EVERY tl.load/tl.store MUST have a mask= argument.
CRITICAL -- DO NOT REPLACE CUBLAS/CUDNN WITH TRITON:
When the ATen ops include matmul (mm, addmm, bmm), linear, or
convolution, Inductor calls cuBLAS/cuDNN for these via extern_kernels.
These library calls are FASTER than any hand-written Triton matmul or
convolution. Do NOT rewrite them as tl.dot loops.
Instead, structure your kernel as:
1. KEEP the matmul/conv as a standard PyTorch call
2. FUSE ONLY the epilogue operations (activation, normalization,
scaling) into a Triton kernel that reads the matmul/conv output
and applies the epilogue in one pass.
BEATING TORCH.COMPILE (INDUCTOR):
1. EPILOGUE FUSION: Fuse bias/activation/normalization after cuBLAS
2. CROSS-REDUCTION FUSION: Single-pass LayerNorm instead of 3 kernels
3. MULTI-OP SEQUENCES: Keep intermediates in registers/SRAM


4. CUSTOM ALGORITHMS: Flash attention, online softmax
5. SPECIALIZED TILING: Hand-tuned @triton.autotune configs
ANTI-CHEATING CONSTRAINTS:
All core computation logic MUST be in @triton.jit kernels.
Banned: torch.matmul, torch.mm, F.linear, F.conv2d, F.layer_norm,
F.gelu, F.softmax, F.scaled_dot_product_attention, extern_kernels.*,
trivial identity/no-op computation.
[Full code examples and fusion patterns omitted for brevity.]
\end{lstlisting}

\subsection{Fusion Agent}
The Fusion Agent fuses adjacent Triton kernels into a single kernel to
eliminate intermediate memory round-trips. It receives the source of $N$
adjacent kernels connected by intermediate buffers and produces a single
fused kernel that keeps intermediate values in registers.
\begin{lstlisting}[style=prompt]
You are an expert GPU kernel engineer specializing in Triton kernel
fusion.
Your task: given N adjacent Triton kernels that are connected by
intermediate buffers, produce a SINGLE fused Triton kernel that
eliminates those buffers and performs all computation in one pass.
Rules:
1. The fused kernel must accept the same external inputs and produce
the same external outputs as the original kernel sequence.
2. Eliminate ALL intermediate buffers listed under
"eliminated_buffers".
3. Preserve numerical correctness: use the same dtypes, same
element-wise computation, and avoid precision loss.
4. The fused kernel must be a valid Triton kernel with @triton.jit.
5. Return ONLY the fused kernel Python source.
6. The function name must be exactly: fused_kernel
7. Include a call wrapper named fused_kernel_call(...).
8. ALL computation must be in the @triton.jit kernel. Do NOT call
torch.nn.*, torch.nn.functional.*, torch.matmul, or any PyTorch
compute API in the wrapper.
FUSION TECHNIQUES:
POINTWISE + POINTWISE: Merge computation bodies. Load inputs once,
apply both operations, store once. Keep intermediates in registers.
REDUCTION + POINTWISE: After the reduction (e.g., mean/variance),
immediately apply the pointwise operation before storing.
NORMALIZATION FUSION (LayerNorm/RMSNorm sequences):
- Compute mean and variance in a single pass
- Normalize, scale, and bias in the same kernel
- Accumulate in fp32 for numerical stability
ACTIVATION + BIAS FUSION: Load data once, add bias, apply activation
(GELU/SiLU/ReLU), store once.
RESIDUAL + NORM: Fuse x = residual + dropout(x); x = LayerNorm(x).
Both operations touch the same data -- one kernel, one memory pass.
\end{lstlisting}

\section{Verification Cascade Details}
\label{app:verification}

Table~\ref{tab:gates} summarizes synthesis validation and the checks
used during search and model evaluation. Section~\ref{sec:correctness}
describes the fallback paths.

\begin{table}[h]
\centering
\footnotesize
\setlength{\tabcolsep}{4pt}
\resizebox{\textwidth}{!}{%
\begin{tabular}{@{}llllll@{}}
\toprule
\textbf{Stage} & \textbf{Gate} & \textbf{Tolerance} & \textbf{Seeds}
  & \textbf{Reference} & \textbf{Scope} \\
\midrule
Synthesis & Eager validation      & $r{=}a{=}10^{-4}$          & 2: $(0,42)$       & PyTorch eager model  & Per synth.\ kernel \\
Gate 1    & Static validation     & N/A (crash check)          & N/A               & N/A                  & Per candidate \\
Gate 2    & Multi-seed corr.      & $10^{-3}$ or generated spec. & 3: $[0,42,1337]$  & Standalone / generated CPU ref. & Per candidate \\
Gate 3    & Model-level (Tier 1)  & $r{=}a{=}10^{-4}$          & 3: $[0,42,123]$   & PyTorch eager model & End-to-end \\
          & Model-level (Tier 2) & Ratio, absolute, or scale-relative bound & Same as Tier 1 & Float64 reference & Model output \\
Gate 4    & Swap performance     & $\gamma=1.03$    & --             & Current assembled model & Per swap \\
\bottomrule
\end{tabular}%
}
\caption{Verification stages. Gates~1--2 screen candidates during
optimization; execution failures can stop testing before Gate~2. Gates~3--4
concern assembled models, with targeted retries on the full-artifact
route. Section~\ref{sec:correctness} describes the numerical and sanity
checks. $r$ and $a$ denote \texttt{rtol} and \texttt{atol}.}
\label{tab:gates}
\end{table}

\paragraph{Weight extraction (Gate 3).}
When Inductor externalizes model parameters as explicit function
arguments, the optimized kernel's input generator would produce random
values for weights. The system intercepts Inductor's argument flattening
via $\texttt{torch.compile}(m, \text{backend}=f_{\text{capture}})$, where
$f_{\text{capture}}$ records the flat arguments observed by the capture
backend. If the argument count matches the candidate interface, the verifier
uses shape and dtype to match captured arguments to original data inputs
and substitutes the current trial's values. Arguments with identical
shape and dtype can be ambiguous under this heuristic.

\section{Shipped Optimization Guidelines}
\label{app:guidelines}
The repository contains benchmark-derived optimization-memory guidelines. Their strategy
labels and attempt counts are not controlled estimates of component effects or
model-level speedup contributions. They are not used as evidence for the
headline aggregate, nor as a substitute for a budget-matched baseline.

\section{Ablation Results}
\label{app:ablation}

We compare the full-system outcomes (C0) with five flag variants
(C1--C5) on a shared 50-problem sample. Table~\ref{tab:ablation} gives
totals, Table~\ref{tab:ablconfig} defines the variants, and
Table~\ref{tab:ablresults} gives the per-level counts.

\begin{table}[h]
\centering
\footnotesize
\setlength{\tabcolsep}{8pt}
\renewcommand{\arraystretch}{1.1}
\begin{tabular}{@{}lrrrr@{}}
\toprule
Config & Opt & Match & Fall & Opt\% \\
\midrule
Full system            & 19 & 14 & 17 & 38\% \\
\midrule
Beam width $B{=}1$      &  5 &  3 & 42 & 10\% \\
Skip-NCU flag          &  5 &  4 & 41 & 10\% \\
Nominal no-memory flag &  4 &  3 & 43 &  8\% \\
No multi-iter ($T{=}1$)&  4 &  4 & 42 &  8\% \\
Per-swap gate off      &  3 &  3 & 44 &  6\% \\
\bottomrule
\end{tabular}
\caption{Outcomes on the same 50 problems for each configuration.
Counts sum the per-level results in Table~\ref{tab:ablresults}.}
\label{tab:ablation}
\end{table}

\begin{table}[h]
\centering
\footnotesize
\begin{tabular}{@{}lll@{}}
\toprule
\textbf{Config} & \textbf{Configuration} & \textbf{Flag} \\
\midrule
C0 & Full system (baseline)   & Reference settings \\
C1 & Beam width $B{=}1$       & \texttt{-{}-beam-width 1} \\
C2 & Nominal no-memory flag   & \texttt{-{}-no-memory} \\
C3 & Per-swap gate off        & \texttt{-{}-no-e2e-perf-gate} \\
C4 & Skip-NCU flag            & \texttt{-{}-skip-ncu} \\
C5 & Single iteration ($T{=}1$) & \texttt{-T 1} \\
\bottomrule
\end{tabular}
\caption{Ablation configurations. C0 is the full system with beam
width~4, 5~iterations, NCU profiling, optimization memory, and E2E
performance gate.}
\label{tab:ablconfig}
\end{table}

\begin{table}[h]
\centering
\footnotesize
\setlength{\tabcolsep}{5pt}
\begin{tabular}{@{}lrrrrrrrrrr@{}}
\toprule
& \multicolumn{3}{c}{\textbf{L1 (20)}}
& \multicolumn{3}{c}{\textbf{L2 (20)}}
& \multicolumn{3}{c}{\textbf{L3 (10)}} & \\
\cmidrule(lr){2-4}\cmidrule(lr){5-7}\cmidrule(lr){8-10}
\textbf{Config} & Opt & Match & Fall & Opt & Match & Fall & Opt & Match & Fall & \textbf{Geomean} \\
\midrule
C0 (full)         & 10 & 6 & 4  & 6 & 7 & 7  & 3 & 1 & 6  & 3.37$\times$ \\
C1 ($B{=}1$)      & 4  & 3 & 13 & 1 & 0 & 19 & 0 & 0 & 10 & -- \\
C2 (nominal flag)& 4  & 3 & 13 & 0 & 0 & 20 & 0 & 0 & 10 & -- \\
C3 (gate off) & 3  & 3 & 14 & 0 & 0 & 20 & 0 & 0 & 10 & -- \\
C4 (skip NCU)       & 3  & 4 & 13 & 2 & 0 & 18 & 0 & 0 & 10 & -- \\
C5 ($T{=}1$)      & 3  & 3 & 14 & 1 & 1 & 18 & 0 & 0 & 10 & -- \\
\bottomrule
\end{tabular}
\caption{Single-run outcomes by level. Categories follow
Table~\ref{tab:results}. The C0 geometric mean covers its 19 optimized
cases. The released artifacts do not include per-problem timings for C1--C5.}
\label{tab:ablresults}
\end{table}

\paragraph{Sampled kernels.}
Before examining full-system outcomes, we drew 50 problems by weighted
random sampling within each benchmark level (20 L1, 20 L2, and 10 L3).
On this sample, the full system optimized 19 problems, matched the
baseline on 14, and fell into the fallback category on 17. The optimized
share was 38\% in the sample and 28\% across the full suite.
\begin{itemize}
\item \textbf{L1 (20):} 002, 003, 006, 007, 009, 012, 015, 022, 028, 036,
  039, 042, 043, 049, 059, 064, 075, 086, 094, 095.
\item \textbf{L2 (20):} 001, 003, 009, 010, 018, 020, 029, 032, 039, 042,
  049, 050, 053, 055, 056, 068, 077, 081, 093, 098.
\item \textbf{L3 (10):} 002, 005, 009, 012, 014, 019, 030, 038, 044, 047.
\end{itemize}

\paragraph{Interpretation.}
C3 yielded 3 optimized cases, compared with 19 for the full system.
The C3 flag disables performance checks and performance-triggered
retries for individual swaps. Correctness checks on swaps and the final
model benchmark still apply. With one run per variant, the comparison
cannot separate the effect of the flag from search variability.

\paragraph{Flag semantics.}
C2 sets the experience queue capacity to zero. The append helper then
evaluates \texttt{memory[-capacity:]} as \texttt{memory[0:]}, which keeps
every item. The planner can still draw on other memory sources.
The \texttt{-{}-optimize-all-kernels} flag (C6) sets
\texttt{needs\_triton\_replacement=True} on traced groups. Group merging
and code generation ignore that field. C4 requests a search without NCU.
In the released implementation, that branch leaves the baseline latency
field unset, although later stages require it.

\section{Isolated Profiling and Model-Level Case Studies}
\label{app:ncu}
Table~\ref{tab:traces} reports NCU durations before and after isolated
kernel optimization, alongside end-to-end speedups over
\texttt{torch.compile}. The isolated durations time standalone kernels,
while the end-to-end speedups compare complete models.

\begin{table}[h]
\centering
\begin{tabular}{lrrr}
\toprule
Problem & Isolated baseline (ns) & Isolated best (ns) & E2E factor\\
\midrule
L1-052 & 59,030,000 & 96,740 & 1.01$\times$\\
L2-034 & 717,410 & 23,330 & 0.994$\times$\\
\bottomrule
\end{tabular}
\caption{L1-052 and L2-034 show isolated speedups of 610.19$\times$ and
30.75$\times$, respectively. Both end-to-end speedups fall in the
matched band.}
\label{tab:traces}
\end{table}
For L1-007, the saved final source casts operands to FP16 before
\texttt{tl.dot} and uses FP32 accumulation. Its 2.89$\times$ E2E result
includes precision and tuning changes together. There is no controlled
measurement assigning separate speedup factors to occupancy, tile size and
precision. A list of autotune configurations is not a record of the setting
actually selected at each execution.

\section{Complete KernelBench Results}
\label{app:perkernel}

Table~\ref{tab:perkernel} presents the complete per-kernel results across
all 250~KernelBench problems (100~L1 + 100~L2 + 50~L3). Category codes:
OPT = optimized (printed speedup ${>}1.03\times$), MAT = matched (kept printed speedup in the $[0.97,1.03]$ reporting band), FALL = fallback-labelled outcome (assigned baseline in aggregate reporting). The Saved column identifies a compiler-wrapper marker (Compile) or generated
source (Generated), without asserting runtime equivalence. Reasons are PR
(performance regression), VF (verification failure, including timeout), and NT
(no generated kernels). A dagger marks multiple terminal report blocks;
a double dagger marks a fallback-labelled run whose saved model is generated
source rather than a compiler wrapper.

{\scriptsize
\setlength{\tabcolsep}{3.5pt}
\begin{longtable}{@{}llrll l !{\vrule width 0.6pt} llrll l@{}}
\caption{Complete per-kernel KernelBench results across all 250~problems,
laid out in two side-by-side blocks (read left block top-to-bottom, then
right block). Speed is terminal logged E2E speedup vs.\ \texttt{torch.compile};
``--'' denotes a fallback-labelled case assigned baseline speedup.}
\label{tab:perkernel}\\
\toprule
\textbf{Lvl} & \textbf{K\#} & \textbf{Speed} & \textbf{Cat} & \textbf{Saved} & \textbf{Rsn}
& \textbf{Lvl} & \textbf{K\#} & \textbf{Speed} & \textbf{Cat} & \textbf{Saved} & \textbf{Rsn} \\
\midrule
\endfirsthead
\multicolumn{12}{@{}l}{\scriptsize\emph{(Table~\ref{tab:perkernel} continued)}}\\
\toprule
\textbf{Lvl} & \textbf{K\#} & \textbf{Speed} & \textbf{Cat} & \textbf{Saved} & \textbf{Rsn}
& \textbf{Lvl} & \textbf{K\#} & \textbf{Speed} & \textbf{Cat} & \textbf{Saved} & \textbf{Rsn} \\
\midrule
\endhead
\midrule
\multicolumn{12}{r@{}}{\scriptsize\emph{Continued on next page}}\\
\endfoot
\bottomrule
\endlastfoot
L1 & 001 & 5.77$\times$ & OPT & Matmul & TO & L1 & 002 & 6.06$\times$ & OPT & Matmul &  \\
L1 & 003 & 5.11$\times$ & OPT & Matmul &  & L1 & 004 & 1.30$\times$ & OPT & Matmul &  \\
L1 & 005 & 1.01$\times$ & OPT &  &  & L1 & 006 & 0.98$\times$ & MAT & Matmul &  \\
L1 & 007 & 2.89$\times$ & OPT & Matmul &  & L1 & 008 & 2.63$\times$ & OPT & Matmul & TO \\
L1 & 009 & 2.07$\times$ & OPT & Matmul &  & L1 & 010 & 5.79$\times$ & OPT & Matmul &  \\
L1 & 011 & 1.05$\times$ & OPT & Einsum &  & L1 & 012 & 88.63$\times$ & OPT & Matmul &  \\
L1 & 013 & 5.79$\times$ & OPT & Matmul &  & L1 & 014 & 14.04$\times$ & OPT & Matmul &  \\
L1 & 015 & 20.26$\times$ & OPT & Matmul &  & L1 & 016 & -- & FALL & Matmul & NI \\
L1 & 017 & 8.65$\times$ & OPT & Matmul & TO & L1 & 018 & 3.03$\times$ & OPT & Matmul &  \\
L1 & 019 & 1.00$\times$ & MAT &  &  & L1 & 020 & 1.01$\times$ & OPT &  &  \\
L1 & 021 & 1.01$\times$ & OPT &  &  & L1 & 022 & 0.97$\times$ & MAT &  &  \\
L1 & 023 & 1.02$\times$ & OPT &  &  & L1 & 024 & 0.97$\times$ & MAT &  &  \\
L1 & 025 & 1.01$\times$ & OPT &  &  & L1 & 026 & 1.05$\times$ & OPT &  &  \\
L1 & 027 & 0.97$\times$ & MAT &  &  & L1 & 028 & 0.97$\times$ & MAT &  &  \\
L1 & 029 & 1.02$\times$ & OPT &  &  & L1 & 030 & 0.98$\times$ & MAT &  &  \\
L1 & 031 & 1.00$\times$ & MAT &  &  & L1 & 032 & 0.97$\times$ & MAT &  &  \\
L1 & 033 & 1.00$\times$ & MAT & BatchNorm &  & L1 & 034 & 1.01$\times$ & OPT & InstanceNorm &  \\
L1 & 035 & -- & FALL & GroupNorm & NI & L1 & 036 & 1.10$\times$ & OPT &  & TO \\
L1 & 037 & 1.01$\times$ & OPT &  &  & L1 & 038 & 1.26$\times$ & OPT &  &  \\
L1 & 039 & 1.38$\times$ & OPT &  &  & L1 & 040 & 1.02$\times$ & OPT & LayerNorm & TO \\
L1 & 041 & -- & OPT & Pooling & TO & L1 & 042 & -- & FALL & Pooling & NI \\
L1 & 043 & 3.41$\times$ & OPT & Pooling & TO & L1 & 044 & -- & FALL & Pooling & NI \\
L1 & 045 & 2.16$\times$ & OPT & Pooling &  & L1 & 046 & 1.00$\times$ & MAT & Pooling &  \\
L1 & 047 & -- & FALL & Reduction & NI & L1 & 048 & 1.22$\times$ & OPT &  &  \\
L1 & 049 & -- & FALL &  & NI & L1 & 050 & 1.02$\times$ & OPT & Conv &  \\
L1 & 051 & -- & FALL &  & NI & L1 & 052 & 1.01$\times$ & OPT &  & TO \\
L1 & 053 & 1.02$\times$ & OPT &  &  & L1 & 054 & 1.00$\times$ & MAT & Conv &  \\
L1 & 055 & 1.06$\times$ & OPT & Conv &  & L1 & 056 & 1.00$\times$ & MAT & Conv &  \\
L1 & 057 & 1.05$\times$ & OPT & ConvTranspose &  & L1 & 058 & -- & FALL & ConvTranspose & NI \\
L1 & 059 & 1.00$\times$ & MAT & Conv &  & L1 & 060 & 1.00$\times$ & MAT & Conv &  \\
L1 & 061 & 1.00$\times$ & MAT & ConvTranspose &  & L1 & 062 & 1.00$\times$ & MAT & Conv &  \\
L1 & 063 & 1.07$\times$ & OPT & Conv &  & L1 & 064 & -- & FALL & ConvTranspose & NI \\
L1 & 065 & 1.03$\times$ & OPT & ConvTranspose &  & L1 & 066 & 1.00$\times$ & MAT & Conv &  \\
L1 & 067 & -- & FALL & Conv & NI & L1 & 068 & 1.00$\times$ & MAT & ConvTranspose &  \\
L1 & 069 & 1.00$\times$ & MAT & ConvTranspose &  & L1 & 070 & 0.99$\times$ & MAT & ConvTranspose &  \\
L1 & 071 & 1.06$\times$ & OPT & ConvTranspose &  & L1 & 072 & 1.00$\times$ & MAT & ConvTranspose &  \\
L1 & 073 & 1.00$\times$ & MAT & ConvTranspose &  & L1 & 074 & -- & FALL & ConvTranspose & NI \\
L1 & 075 & 1.00$\times$ & MAT & ConvTranspose &  & L1 & 076 & -- & FALL & Conv & NI \\
L1 & 077 & 1.00$\times$ & MAT & ConvTranspose &  & L1 & 078 & 1.00$\times$ & MAT & ConvTranspose &  \\
L1 & 079 & 1.00$\times$ & MAT & ConvTranspose &  & L1 & 080 & -- & FALL & Conv & NI \\
L1 & 081 & 1.03$\times$ & OPT & ConvTranspose &  & L1 & 082 & 1.05$\times$ & OPT & Conv &  \\
L1 & 083 & 1.00$\times$ & MAT & Conv &  & L1 & 084 & 1.06$\times$ & OPT & Conv &  \\
L1 & 085 & 11.38$\times$ & OPT & Conv &  & L1 & 086 & -- & FALL & Conv & NI \\
L1 & 087 & 1.32$\times$ & OPT & Conv &  & L1 & 088 & 1.00$\times$ & MAT &  &  \\
L1 & 089 & 1.14$\times$ & OPT & CumSum &  & L1 & 090 & 1.19$\times$ & OPT &  &  \\
L1 & 091 & 2.80$\times$ & OPT & CumSum &  & L1 & 092 & 1.65$\times$ & OPT & CumSum, Cat &  \\
L1 & 093 & 1.08$\times$ & OPT & CumSum &  & L1 & 094 & 1.00$\times$ & MAT &  &  \\
L1 & 095 & 1.70$\times$ & OPT &  &  & L1 & 096 & 1.00$\times$ & MAT &  &  \\
L1 & 097 & -- & FALL &  & NI & L1 & 098 & 1.00$\times$ & MAT &  & TO \\
L1 & 099 & 1.03$\times$ & OPT &  &  & L1 & 100 & 1.00$\times$ & MAT &  &  \\
L2 & 001 & 1.00$\times$ & MAT & Conv &  & L2 & 002 & 1.00$\times$ & MAT & ConvTranspose &  \\
L2 & 003 & 1.00$\times$ & MAT & LN, Pool, GELU & TO & L2 & 004 & 1.06$\times$ & OPT & Conv &  \\
L2 & 005 & 1.01$\times$ & OPT & ConvTranspose &  & L2 & 006 & 1.03$\times$ & OPT & Conv, Pool &  \\
L2 & 007 & 1.00$\times$ & MAT & Conv &  & L2 & 008 & 1.19$\times$ & OPT & Conv, Red, Pool &  \\
L2 & 009 & 1.47$\times$ & OPT & Linear/GEMM &  & L2 & 010 & -- & FALL & Pool, ConvT & C \\
L2 & 011 & -- & FALL & BN, Pool, ConvT & C & L2 & 012 & -- & FALL & Linear/GEMM & G \\
L2 & 013 & 1.00$\times$ & MAT & ConvTranspose & TO & L2 & 014 & -- & FALL & Matmul, Red & G \\
L2 & 015 & 1.00$\times$ & MAT & BN, ConvT &  & L2 & 016 & 1.06$\times$ & OPT & ConvTranspose & TO \\
L2 & 017 & 0.98$\times$ & MAT & Conv, IN & TO & L2 & 018 & 33.11$\times$ & OPT & Linear, Red &  \\
L2 & 019 & 1.09$\times$ & OPT & ConvT, GN & TO & L2 & 020 & 1.00$\times$ & MAT & ConvTranspose &  \\
L2 & 021 & 0.98$\times$ & MAT & Conv, GN &  & L2 & 022 & 7.18$\times$ & OPT & Linear/GEMM &  \\
L2 & 023 & 1.03$\times$ & OPT & Conv, GN & TO & L2 & 024 & 1.06$\times$ & OPT & Conv &  \\
L2 & 025 & 1.23$\times$ & OPT & Conv & TO & L2 & 026 & 1.00$\times$ & MAT & ConvTranspose &  \\
L2 & 027 & 1.00$\times$ & MAT & Conv, GN & TO & L2 & 028 & -- & FALL & Linear, IN & G \\
L2 & 029 & -- & FALL & Linear/GEMM & G & L2 & 030 & -- & FALL & Linear, GN & G \\
L2 & 031 & 1.00$\times$ & MAT & Conv &  & L2 & 032 & 1.04$\times$ & OPT & Conv & TO \\
L2 & 033 & -- & FALL & Linear, BN & G & L2 & 034 & 0.99$\times$ & MAT & LN, ConvT &  \\
L2 & 035 & -- & FALL & Conv, Pool & C & L2 & 036 & 1.04$\times$ & OPT & Red, ConvT &  \\
L2 & 037 & -- & FALL & Linear, GN & G & L2 & 038 & 1.03$\times$ & OPT & Pool, ConvT &  \\
L2 & 039 & -- & FALL & Linear, BN & G & L2 & 040 & 1.51$\times$ & OPT & Linear/GEMM &  \\
L2 & 041 & -- & FALL & Linear, BN & G & L2 & 042 & 0.98$\times$ & MAT & Red, ConvT & TO \\
L2 & 043 & 1.00$\times$ & MAT & Conv, Pool &  & L2 & 044 & 0.99$\times$ & MAT & ConvTranspose & TO \\
L2 & 045 & -- & FALL & Linear/GEMM & G & L2 & 046 & -- & FALL & Conv, Pool & C \\
L2 & 047 & 0.97$\times$ & MAT & Conv &  & L2 & 048 & 1.00$\times$ & MAT & Conv &  \\
L2 & 049 & 1.17$\times$ & OPT & ConvTranspose &  & L2 & 050 & 1.00$\times$ & MAT & Pool, ConvT &  \\
L2 & 051 & -- & FALL & Linear/GEMM & G & L2 & 052 & 1.00$\times$ & MAT & Conv, BN &  \\
L2 & 053 & 1.60$\times$ & OPT & Linear, GELU &  & L2 & 054 & 1.00$\times$ & MAT & Conv & TO \\
L2 & 055 & 5.46$\times$ & OPT & Linear, Red, Pool &  & L2 & 056 & -- & FALL & Linear, Red & G \\
L2 & 057 & 1.00$\times$ & MAT & Conv &  & L2 & 058 & 1.00$\times$ & MAT & ConvTranspose &  \\
L2 & 059 & -- & FALL & Linear/GEMM & G & L2 & 060 & 1.01$\times$ & OPT & ConvT, GN &  \\
L2 & 061 & 1.02$\times$ & OPT & ReLU, ConvT, GN &  & L2 & 062 & -- & FALL & Linear, GN & G \\
L2 & 063 & -- & FALL & Linear/GEMM & G & L2 & 064 & 8.30$\times$ & OPT & Linear/GEMM &  \\
L2 & 065 & 1.00$\times$ & MAT & Conv, Red, Pool & TO & L2 & 066 & -- & FALL & Linear/GEMM & G \\
L2 & 067 & -- & FALL & Conv & C & L2 & 068 & -- & FALL & Linear/GEMM & G \\
L2 & 069 & 1.00$\times$ & MAT & Conv &  & L2 & 070 & 1.43$\times$ & OPT & Linear/GEMM &  \\
L2 & 071 & -- & FALL & Conv & C & L2 & 072 & 1.00$\times$ & MAT & BN, Pool, ConvT &  \\
L2 & 073 & 1.03$\times$ & OPT & Conv, BN &  & L2 & 074 & 1.00$\times$ & MAT & Pool, ConvT &  \\
L2 & 075 & -- & FALL & Linear, GN & G & L2 & 076 & -- & FALL & Linear/GEMM & G \\
L2 & 077 & 1.00$\times$ & MAT & BN, ConvT & TO & L2 & 078 & 1.00$\times$ & MAT & Red, Pool, ConvT &  \\
L2 & 079 & 1.01$\times$ & OPT & Conv, IN & TO & L2 & 080 & 5.78$\times$ & OPT & Linear/GEMM &  \\
L2 & 081 & -- & FALL & Linear/GEMM & G & L2 & 082 & 1.00$\times$ & MAT & Conv, Pool & TO \\
L2 & 083 & 1.00$\times$ & MAT & Conv, GN &  & L2 & 084 & -- & FALL & Linear, BN & G \\
L2 & 085 & 1.10$\times$ & OPT & Conv, Pool, GN &  & L2 & 086 & -- & FALL & Linear/GEMM & G \\
L2 & 087 & 1.17$\times$ & OPT & Conv &  & L2 & 088 & -- & FALL & Linear, GN & G \\
L2 & 089 & 1.00$\times$ & MAT & Pool, ConvT &  & L2 & 090 & 1.00$\times$ & MAT & Conv &  \\
L2 & 091 & 1.04$\times$ & OPT & ConvTranspose & TO & L2 & 092 & 1.06$\times$ & OPT & Conv, GN & TO \\
L2 & 093 & 1.00$\times$ & MAT & ConvTranspose &  & L2 & 094 & -- & FALL & Linear, GN & G \\
L2 & 095 & -- & FALL & Linear/GEMM & G & L2 & 096 & 1.10$\times$ & OPT & Pool, ConvT & TO \\
L2 & 097 & -- & FALL & Linear, BN & G & L2 & 098 & -- & FALL & Linear, Pool & G \\
L2 & 099 & -- & FALL & Linear/GEMM & G & L2 & 100 & 1.00$\times$ & MAT & ConvTranspose & TO \\
L3 & 001 & -- & FALL & Linear, ReLU & G & L3 & 002 & 4.12$\times$ & OPT & Linear, ReLU &  \\
L3 & 003 & -- & FALL & Linear, ReLU & CE & L3 & 004 & 0.97$\times$ & MAT & Conv, Linear, ReLU &  \\
L3 & 005 & -- & FALL & Conv, Linear, ReLU & C & L3 & 006 & -- & FALL & Conv, Pool, Cat & C \\
L3 & 007 & -- & FALL & Conv, Linear, ReLU & CE & L3 & 008 & -- & FALL & Conv, BN, ReLU & C \\
L3 & 009 & 1.55$\times$ & OPT & Conv, Linear, BN &  & L3 & 010 & -- & FALL & Conv, Linear, BN & C \\
L3 & 011 & -- & FALL & Conv, Linear, ReLU & C & L3 & 012 & -- & FALL & Conv, Linear, ReLU & C \\
L3 & 013 & -- & FALL & Conv, BN, ReLU & C & L3 & 014 & 0.98$\times$ & MAT & Conv, BN, ReLU & TO \\
L3 & 015 & 1.15$\times$ & OPT & Conv, Linear, BN &  & L3 & 016 & 1.06$\times$ & OPT & Conv, Linear, BN &  \\
L3 & 017 & -- & FALL & Conv, ReLU, Cat & C & L3 & 018 & -- & FALL & Conv, ReLU, Pool & C \\
L3 & 019 & -- & FALL & Conv, Linear, BN & C & L3 & 020 & 1.03$\times$ & OPT & Conv, Linear, BN &  \\
L3 & 021 & -- & FALL & Conv, BN, ReLU & C & L3 & 022 & -- & FALL & Conv, Linear, BN & C \\
L3 & 023 & -- & FALL & Conv, Linear, BN & C & L3 & 024 & 2.33$\times$ & OPT & Conv, Linear, BN & TO \\
L3 & 025 & -- & FALL & Conv, BN, ReLU & C & L3 & 026 & -- & FALL & Conv, Linear, BN & C \\
L3 & 027 & 1.00$\times$ & MAT & Conv, Linear, BN &  & L3 & 028 & -- & FALL & Linear, GELU, Cat & G \\
L3 & 029 & -- & FALL & Conv, Linear, LN & C & L3 & 030 & -- & FALL & Conv, Linear, LN & C \\
L3 & 031 & 1.10$\times$ & OPT & LN, Attention & TO & L3 & 032 & 1.11$\times$ & OPT & Conv, Linear, Cat &  \\
L3 & 033 & -- & FALL & Linear, Cat & CE & L3 & 034 & -- & FALL & Linear, Cat & G \\
L3 & 035 & -- & FALL & Linear, LSTM & SF & L3 & 036 & -- & FALL & Linear, LSTM & NT \\
L3 & 037 & -- & FALL & Linear, LSTM & NT & L3 & 038 & -- & FALL & Linear, LSTM & NT \\
L3 & 039 & -- & FALL & GRU & NT & L3 & 040 & -- & FALL & GRU & NT \\
L3 & 041 & -- & FALL & GRU & NT & L3 & 042 & -- & FALL & GRU & NT \\
L3 & 043 & -- & FALL & Linear, Attn, MM & G & L3 & 044 & -- & FALL & Linear, LN, Attn & G \\
L3 & 045 & 1.06$\times$ & OPT & Conv, BN, Pool & TO & L3 & 046 & 1.09$\times$ & OPT & BN, Softmax, Red &  \\
L3 & 047 & 1.08$\times$ & OPT & BN, Softmax, Red & TO & L3 & 048 & 1.17$\times$ & OPT & CumSum, Cat, Ein &  \\
L3 & 049 & -- & FALL & CumSum, Cat, Ein & PR & L3 & 050 & -- & FALL & Linear, Attn, MM & G \\

\end{longtable}
}

\section{Generated Kernel Examples}
\label{app:examples}

Below are saved optimized models and task-level algebraic explanations.
They do not isolate the contribution of individual search components.

\subsection{L1 Kernel 012: Diagonal Matrix Multiply (88.63$\times$)}
The benchmark expression constructs a dense diagonal matrix and multiplies it
by another matrix. The saved optimized model uses element-wise row scaling,
reducing arithmetic from a dense $O(N^3)$ product to $O(N^2)$ work for the
benchmark dimensions. The 88.63$\times$ E2E factor is specific to this
structured task and does not measure generic dense-GEMM acceleration or isolate
an NCU-guided optimization contribution.
\begin{lstlisting}[style=pycode]
@triton.autotune(
    configs=[
        triton.Config({'BLOCK_ROWS': 1, 'BLOCK_COLS': 1024}, num_warps=4),
        triton.Config({'BLOCK_ROWS': 1, 'BLOCK_COLS': 512},  num_warps=4),
        triton.Config({'BLOCK_ROWS': 2, 'BLOCK_COLS': 512},  num_warps=4),
        triton.Config({'BLOCK_ROWS': 4, 'BLOCK_COLS': 256},  num_warps=4),
        triton.Config({'BLOCK_ROWS': 1, 'BLOCK_COLS': 256},  num_warps=2),
        triton.Config({'BLOCK_ROWS': 8, 'BLOCK_COLS': 128},  num_warps=4),
        triton.Config({'BLOCK_ROWS': 1, 'BLOCK_COLS': 2048}, num_warps=8),
        triton.Config({'BLOCK_ROWS': 2, 'BLOCK_COLS': 1024}, num_warps=8),
    ],
    key=['num_rows', 'num_cols'],
)
@triton.jit
def diag_matmul_kernel(
    a_ptr, b_ptr, out_ptr,
    num_rows, num_cols,
    stride_b_row, stride_b_col,
    stride_out_row, stride_out_col,
    BLOCK_ROWS: tl.constexpr,
    BLOCK_COLS: tl.constexpr,
):
    pid_row = tl.program_id(0)
    pid_col = tl.program_id(1)
    row_start = pid_row * BLOCK_ROWS
    col_start = pid_col * BLOCK_COLS
    row_offsets = row_start + tl.arange(0, BLOCK_ROWS)
    col_offsets = col_start + tl.arange(0, BLOCK_COLS)
    row_mask = row_offsets < num_rows
    col_mask = col_offsets < num_cols
    # Load diagonal elements A[row]
    a_vals = tl.load(a_ptr + row_offsets, mask=row_mask, other=0.0)
    # Load B block [BLOCK_ROWS, BLOCK_COLS]
    b_ptrs = (b_ptr
              + row_offsets[:, None] * stride_b_row
              + col_offsets[None, :] * stride_b_col)
    full_mask = row_mask[:, None] & col_mask[None, :]
    b_vals = tl.load(b_ptrs, mask=full_mask, other=0.0)
    # Scale each row by diagonal element: O(N^2)
    out_vals = a_vals[:, None] * b_vals
    out_ptrs = (out_ptr
                + row_offsets[:, None] * stride_out_row
                + col_offsets[None, :] * stride_out_col)
    tl.store(out_ptrs, out_vals, mask=full_mask)
\end{lstlisting}

\subsection{L2 Kernel 018: Algebraic Simplification (33.11$\times$)}
This kernel computes a 6-operation chain:
Linear~$\to$~Sum~$\to$~Max~$\to$~Mean~$\to$~LogSumExp~$\to$~LogSumExp. The
input is an $(M, K)$ matrix $X$, and the linear layer produces an
$(M, N)$ intermediate via $Y = XW^\top + b$. Subsequent operations (sum,
max, mean, LogSumExp) all reduce dimension~1, collapsing each $(M, N)$ row
to a scalar. KernelOPT's synthesis agent discovers the algebraic
identity:
\begin{equation*}
\textstyle\sum_j \big( \sum_k X_{ik} W_{jk} + b_j \big)
= \sum_k X_{ik} \cdot \big( \sum_j W_{jk} \big) + \sum_j b_j .
\end{equation*}
This reduces the computation from an $(M, K)\times(K, N)$ matrix multiply
(dense matrix product) followed by per-row reductions on $(M, N)$, to a single
matrix--vector product. Since the max, mean, and both LogSumExp operations
act on a singleton dimension after the sum, each reduces to an identity.
The saved \texttt{kernel\_function} computes the weight column sums
$w_k=\sum_j W_{jk}$ and bias sum $b_{\text{sum}}=\sum_j b_j$ on each
invocation before launching the Triton kernel. The end-to-end timed call
includes these reductions. The total arithmetic work is $O(NK+MK)$,
including the weight reduction, versus $O(MNK)$ for the dense product.
\begin{lstlisting}[style=pycode]
@triton.jit
def fused_linear_sum_logsumexp_kernel(
    X_ptr, W_ptr, B_ptr, OUT_ptr,
    M, N, K,
    stride_xm, stride_xk, stride_wn, stride_wk,
    BLOCK_M: tl.constexpr, BLOCK_K: tl.constexpr,
):
    # W_ptr points to precomputed w_col_sum[k] = sum_j(W[j,k])
    # B_ptr points to precomputed b_sum = sum_j(B[j])
    pid = tl.program_id(0)
    rows = pid * BLOCK_M + tl.arange(0, BLOCK_M)
    row_mask = rows < M
    acc = tl.zeros([BLOCK_M], dtype=tl.float32)
    for k_off in range(0, K, BLOCK_K):
        k_idx = k_off + tl.arange(0, BLOCK_K)
        k_mask = k_idx < K
        w_col_sum = tl.load(W_ptr + k_idx, mask=k_mask, other=0.0)
        x_ptrs = (X_ptr + rows[:, None] * stride_xm
                  + k_idx[None, :] * stride_xk)
        x_vals = tl.load(x_ptrs,
                         mask=row_mask[:, None] & k_mask[None, :],
                         other=0.0)
        acc += tl.sum(x_vals * w_col_sum[None, :], axis=1)
    b_sum = tl.load(B_ptr)
    acc += b_sum
    tl.store(OUT_ptr + rows, acc, mask=row_mask)
\end{lstlisting}

\subsection{L1 Kernel 007: Precision-changing matmul (2.89$\times$)}
The saved model casts operands to FP16 before \texttt{tl.dot} with an FP32
accumulator and supplies an autotune configuration list. The reported
2.89$\times$ E2E gain includes these changes together. A configuration list
is not evidence of the configuration actually selected in every measurement;
we do not report a winning launch setting without its measurement record.
Matched-precision baselines and controlled variants are needed to attribute
the gain to occupancy, precision, or tiling.

\section{Matrix-Product Precision Sensitivity}
\label{app:precision}

Table~\ref{tab:precision-cases} identifies 17 optimized matrix-product tasks
with explicit reduced-precision operands, TF32 permission, or unspecified dot
precision. These labels describe source settings, not measured GPU instructions.
In particular, permitting TF32 does not record the instruction chosen, and
L3-002 relies on the compiler default, which can depend on environment settings.
The list is not an exhaustive precision classification of all 250 problems:
convolution comparisons also require the corresponding library precision policy.

\begin{table}[h]
\centering
\footnotesize
\begin{tabular}{lp{0.67\textwidth}}
\toprule
Source setting & Problems\\
\midrule
FP16 operands & L1: 001, 003, 007, 009, 010, 013, 014, 015; L2: 080\\
BF16 operands & L1: 002\\
TF32 permitted & L1: 008, 017, 018; L2: 022, 055, 064\\
Dot precision unspecified & L3: 002\\
\bottomrule
\end{tabular}
\caption{Source-level precision settings in 17 optimized matrix-product cases.}
\label{tab:precision-cases}
\end{table}

Let $\mathcal I_\ell$ be the set of problem indices belonging to level $\ell$,
$n_\ell=|\mathcal I_\ell|$, and $\mathcal P_\ell\subseteq\mathcal I_\ell$
the listed precision cases. Let $s_i$ be the reported factor, with
fallback-labelled cases assigned one. The sensitivity
calculation retains every problem and replaces only the listed factors:
\[
G_\ell^{\rm sens}=\exp\!\left(\frac{1}{n_\ell}
\sum_{i\in\mathcal I_\ell\setminus\mathcal P_\ell}\log s_i\right).
\]
For $n_\ell=100,100,50$, the results are 1.141609, 1.067050 and 1.041600.
The listed cases contribute 57.4\% of $\sum_{i=1}^{250}\log s_i$.
Replacing a factor by one is a sensitivity calculation, not a measured
matched-precision result or an estimate of what removing a precision change
would do. No TF32-enabled-baseline re-timing is available.

\section{Tutorial Kernels}
\label{app:tutorials}

The supplementary evaluation applies isolated kernel search to 10
Triton tutorials and 10 Helion~\citep{helion2025} tutorials. Four Triton
kernels improve, with a maximum of $1.70\times$ for matrix multiplication
using TF32. Five Helion kernels improve, with a maximum of $1.12\times$.
These timings concern standalone kernels and use the isolated checks in
Section~\ref{sec:correctness}.

\section{Gate 3 Correctness Verification: \texorpdfstring{$\rho$}{rho} Distribution Analysis}
\label{app:rho_dist}

The float64 check produced 167 rejection messages across 46 of the
250 KernelBench problems. Six came from the scale-relative test for
near-zero reference error. The remaining 161 report $\rho$, with a
minimum of 1,569.3, a median of 3,910.1, and a maximum of
$4.2794\times10^8$. These counts include repeated messages.

The ratios describe rejected outputs. Equation~\ref{eq:rho} also allows
an output with $\rho>10$ to pass if its absolute error is within the
bound. Estimating false acceptance would require independent checks of
accepted outputs. Estimating retry recovery would require linking each
rejection to the outcome of its retry.

\section{Hardware and Software Configuration}
\label{app:hw}

\begin{table}[h]
\centering
\footnotesize
\begin{tabular}{@{}ll@{}}
\toprule
\textbf{Component} & \textbf{Specification} \\
\midrule
GPU          & NVIDIA H200 SXM (Hopper, CC 9.0) \\
SMs          & 132 \\
HBM3e        & 141\,GB, 4.8\,TB/s bandwidth \\
L2 Cache     & 50\,MB \\
CUDA build (PyTorch) & 13.0 \\
PyTorch      & 2.11.0+cu130 \\
Triton       & 3.6.0 \\
Nsight Compute & 2026.2.1.0 \\
Python       & 3.12.16 \\
LLM          & Claude Sonnet 4.6 (Anthropic) \\
LLM API      & Google Cloud Vertex AI \\
LLM timeout  & 900\,s per call \\
Maximum beam width & 4 chains \\
Iteration limit & 5 per isolated target \\
Plan budget ($N$) & 4 shared across active chains \\
Executor parameter ($K$) & 4 (nested budgets; \S\ref{sec:executor-budget}) \\
E2E retry limit & 3 (full-artifact route) \\
Perf gate $\gamma$ & 1.03 (3\% margin) \\
\bottomrule
\end{tabular}
\caption{Hardware, software, and search settings.}
\label{tab:hwconfig}
\end{table}

\end{document}